\PassOptionsToPackage{hyphens}{url}

\documentclass[times, review, 10pt]{elsarticle}
\usepackage[a4paper, top=4.3cm, bottom=4.3cm, left=4.8cm, right=4.8cm]{geometry}

\usepackage{amssymb}
\usepackage{amsmath}
\usepackage{subcaption}
\usepackage{booktabs}
\usepackage{url}         % basic URL support
\usepackage[hyphens]{url} % allow URL line breaks at hyphens

\usepackage{lineno}

\journal{Pattern Recognition}

\begin{document}

\begin{frontmatter}

%% Title, authors and addresses

%% use the tnoteref command within \title for footnotes;
%% use the tnotetext command for theassociated footnote;
%% use the fnref command within \author or \affiliation for footnotes;
%% use the fntext command for theassociated footnote;
%% use the corref command within \author for corresponding author footnotes;
%% use the cortext command for theassociated footnote;
%% use the ead command for the email address,
%% and the form \ead[url] for the home page:
%% \title{Title\tnoteref{label1}}
%% \tnotetext[label1]{}
%% \author{Name\corref{cor1}\fnref{label2}}
%% \ead{email address}
%% \ead[url]{home page}
%% \fntext[label2]{}
%% \cortext[cor1]{}
%% \affiliation{organization={},
%%             addressline={},
%%             city={},
%%             postcode={},
%%             state={},
%%             country={}}
%% \fntext[label3]{}

\title{Beyond Single-Channel: Multichannel Signal Imaging for PPG-to-ECG Reconstruction with Vision Transformers}

%% use optional labels to link authors explicitly to addresses:
%% \author[label1,label2]{}
%% \affiliation[label1]{organization={},
%%             addressline={},
%%             city={},
%%             postcode={},
%%             state={},
%%             country={}}
%%
%% \affiliation[label2]{organization={},
%%             addressline={},
%%             city={},
%%             postcode={},
%%             state={},
%%             country={}}

\author[inst1]{Xiaoyan Li} %% Author name
\author[inst2]{Shixin Xu}
\author[inst3]{Faisal Habib} 
\author[inst1]{Arvind Gupta}
\author[inst4]{Huaxiong Huang}

%% Author affiliation
\affiliation[inst1]{
  organization={Department of Computer Science, University of Toronto}, 
  addressline={40 St. George Street}, 
  city={Toronto},
  postcode={M5S 1A1}, 
  state={Ontario},
  country={Canada}
}
\affiliation[inst2]{
  organization={Data Science Research Center, Duke Kunshan University}, 
  addressline={No. 8 Duke Avenue}, 
  city={Kunshan},
  postcode={215300}, 
  state={Jiangsu},
  country={China}
}
\affiliation[inst3]{
  organization={Mathematics, Analytics, and Data Science Lab, Fields Institute for Research in Mathematical Sciences}, 
  addressline={222 College Street}, 
  city={Toronto},
  postcode={M5T 3J1}, 
  state={Ontario},
  country={Canada}
}
\affiliation[inst4]{
  organization={Department of Mathematics and Statistics, York University}, 
  addressline={4700 Keele Street}, 
  city={Toronto},
  postcode={M3J 1P3}, 
  state={Ontario},
  country={Canada}
}

%% Abstract
\begin{abstract}
Reconstructing ECG from PPG is a promising yet challenging task. While recent advancements in generative models have significantly improved ECG reconstruction, accurately capturing fine-grained waveform features remains a key challenge. To address this, we propose a novel PPG-to-ECG reconstruction method that leverages a Vision Transformer (ViT) as the core network. Unlike conventional approaches that rely on single-channel PPG, our method employs a four-channel signal image representation, incorporating the original PPG, its first-order difference, second-order difference, and area under the curve. This multi-channel design enriches feature extraction by preserving both temporal and physiological variations within the PPG. By leveraging the self-attention mechanism in ViT, our approach effectively captures both inter-beat and intra-beat dependencies, leading to more robust and accurate ECG reconstruction. Experimental results demonstrate that our method consistently outperforms existing 1D convolution-based approaches, achieving up to 29\% reduction in PRD and 15\% reduction in RMSE. The proposed approach also produces improvements in other evaluation metrics, highlighting its robustness and effectiveness in reconstructing ECG signals. Furthermore, to ensure a clinically relevant evaluation, we introduce new performance metrics, including QRS area error, PR interval error, RT interval error, and RT amplitude difference error. Our findings suggest that integrating a four-channel signal image representation with the self-attention mechanism of ViT enables more effective extraction of informative PPG features and improved modeling of beat-to-beat variations for PPG-to-ECG mapping. Beyond demonstrating the potential of PPG as a viable alternative for heart activity monitoring, our approach opens new avenues for cyclic signal analysis and prediction.
\end{abstract}

%%Graphical abstract
% \begin{graphicalabstract}
%\includegraphics{grabs}
% \end{graphicalabstract}

% \begin{highlights}
% \item Multi-channel PPG representation improves ECG reconstruction accuracy
% \item AUC and signal differences enrich physiologically relevant PPG features
% \item Vision Transformer captures temporal and intra-beat ECG relationships
% \item New clinical metrics enable detailed ECG waveform quality evaluation
% \end{highlights}

%% Keywords
\begin{keyword}
%% keywords here, in the form: keyword \sep keyword

%% PACS codes here, in the form: \PACS code \sep code

%% MSC codes here, in the form: \MSC code \sep code
%% or \MSC[2008] code \sep code (2000 is the default)

PPG-to-ECG reconstruction \sep multi-channel signal representation \sep Vision Transformer \sep ECG image \sep ECG waveform \sep AUC
\end{keyword}

\end{frontmatter}

%% Add \usepackage{lineno} before \begin{document} and uncomment 
%% following line to enable line numbers
% \linenumbers

\section{Introduction}
\label{sec:introduction}
Electrocardiograms (ECGs) are essential tools for diagnosing and monitoring cardiovascular health, providing crucial insights into heart rate variability (HRV), heart rate, and key waveform features. These include the QRS complex, PR interval, ST segment, TP interval, and QT interval, which are vital for understanding the heart's electrical activity and diagnosing various cardiac conditions \cite{Goldberger2017ECG, sornmo2000knowledge}. For example, a normal PR interval ranges from 120 to 200 milliseconds. Prolonged PR intervals may indicate first-degree atrioventricular (AV) block or delayed conduction through the AV node, suggesting potential cardiac conduction issues \cite{surawicz2009}. Conversely, shortened PR intervals might imply conditions such as Wolff-Parkinson-White (WPW) syndrome or Lown-Ganong-Levine syndrome, where accessory pathways bypass the normal AV nodal delay \cite{surawicz2009}.

Similarly, prolonged ST segments are indicative of myocardial ischemia or acute myocardial infarction, resulting from delayed repolarization, or may reflect ventricular conduction defects that affect the timing of repolarization \cite{sornmo2000knowledge}. On the other hand, abnormally short ST segments can be a sign of hyperkalemia, which typically shortens ventricular repolarization, or congenital short QT syndrome, a rare but serious condition associated with a high risk of arrhythmias \cite{surawicz2009}. Building on these diagnostic capabilities, recent research has explored using ECG signals for noninvasive serum electrolyte prediction and monitoring \cite{lin2022point,von2024evaluating}. 

Despite the clinical significance of detailed waveform features, much of the existing research on ECG reconstruction from photoplethysmography (PPG) has primarily focused on estimating HRV and heart rate \cite{Reiss2019sensors, s24010141}. However, these approaches often struggle to accurately recover smaller, yet diagnostically important, waveform components such as the T and P waves \cite{wu2025deep}. To address this limitation, we propose a novel method that emphasizes the precise reconstruction of these critical ECG features. By prioritizing accurate prediction and measurement of complete waveforms, our approach aims to enable more comprehensive evaluation and enhanced diagnostic capability.

Reconstructing ECG signals from PPG addresses several limitations of traditional ECG monitoring, including restricted mobility, skin irritation, and the reliance on offline data processing \cite{tian2023}. As a non-invasive, wearable-compatible technology, PPG offers significant advantages for long-term and real-time cardiovascular monitoring \cite{Nie2024Review}. For instance, Gil et al. \cite{gil2010photoplethysmography} demonstrated the value of PPG in assessing vascular and respiratory functions, while Chua et al. \cite{chua2010towards} explored its application in nocturnal blood pressure estimation. Nevertheless, ECG remains the clinical gold standard due to its superior resolution of key waveform components crucial for cardiac diagnostics. The intrinsic physiological relationship between ECG and PPG, where cardiac electrical activity governs peripheral blood volume changes \cite{El-Hajj2020}, underpins the rationale for PPG-to-ECG reconstruction models.

Drawing from established approaches in ECG analysis, the task of mapping the complex, non-linear relationship between ECG and PPG signals can be framed as a pattern recognition problem, with the objective of learning temporal and morphological correspondences between the two modalities \cite{sornmo2000knowledge}. This task is challenging due to signal artifacts and the inherent physiological differences: ECG captures the heart's electrical activity, while PPG measures peripheral blood volume changes. Recent advances in pattern recognition, especially deep learning for time-series and physiological signal analysis, have shown significant promise in addressing these challenges \cite{KIM20221, DONIDALABATI20231}. 

A variety of deep learning models have been proposed for PPG-to-ECG translation. For example, Zhu et al. \cite{Zhu2021} developed a convolutional neural network (CNN) to reconstruct ECG waveforms from single-channel PPG. Tang et al. \cite{Tang2023} introduced an alignment-based LSTM model emphasizing temporal synchronization between PPG and ECG cycles. Tian et al. \cite{tian2023} proposed a joint dictionary learning framework to model cross-domain correspondences. Chiu et al. \cite{Chiu2020} employed attentional neural networks to reconstruct QRS complexes from PPG input. More recently, Shome et al. \cite{Shome2024} proposed a region-disentangled diffusion model to selectively enhance high-information ECG segments, such as the QRS complex.

However, most existing methods rely on single-channel PPG signals, which limits their capacity to capture waveform variability across cardiac cycles. To overcome this limitation, we propose a multi-channel pattern recognition approach using 2D signal representations. Specifically, we generate a four-channel PPG image by stacking padded PPG cycles and computing three derived features: the first-order difference, the second-order difference, and the area under the curve (AUC). This enriched representation allows the model to capture both local waveform structures and inter-beat variability, thereby improving the accuracy and fidelity of ECG signal reconstruction.

Traditional CNNs have demonstrated strong performance in signal processing tasks, including ECG reconstruction. However, their reliance on localized filters and limited receptive fields constrains their capacity to capture long-range dependencies and global contextual patterns. Recent advances in self-attention-based models, particularly Transformers \cite{vaswani2017attention}, have addressed these limitations by effectively modeling sequential dependencies through global context. These models have shown strong performance across a wide range of time-series applications \cite{YU2024110552, NAYAK2024100716}. Vision Transformers (ViTs) \cite{dosovitskiy2021image}, an extension of the Transformer architecture for image-based analysis, further enhance this capability by capturing spatial structures in data-rich image representations.

Building on these advancements, we propose a novel ECG reconstruction method that applies a ViT architecture to four-channel image-like representations of PPG signals. Unlike conventional 1D CNN-based models, which operate on raw PPG sequences, our method restructures the input as a 3D tensor of size $X \times Y \times Z$: the $X$-$Y$ plane captures a 2D signal image formed by stacking individually padded PPG cycles along temporal and morphological dimensions, while $Z$ denotes the number of feature channels. These include the raw PPG signal, its first-order difference, second-order difference, and the AUC.

This transformation from a 1D sequence to a structured 2D image enables the use of 2D patching in ViT, where each patch captures both local waveform morphology and cross-cycle dynamics. This 2D formulation implicitly incorporates strided segmentation along the temporal axis, similar to applying overlapping or non-overlapping patches in 1D sequence models. The $Z$-dimension further enriches this representation by incorporating derivative and cumulative signal features, improving the model's ability to extract clinically relevant information. Through self-attention, the ViT captures dependencies within and across beats, modeling both intra-beat patterns and inter-beat temporal relationships. As a result, our approach addresses the key limitations of traditional 1D sequence-based models and offers enhanced fidelity in ECG waveform reconstruction.

Our contributions are fourfold:
\begin{enumerate}
    \item We propose a novel four-channel image-based representation for ECG reconstruction, diverging from traditional single-channel PPG inputs. Our formulation combines the raw PPG signal with its first-order difference, second-order difference, and AUC, enriching the signal representation and enabling more comprehensive feature extraction.

    \item By incorporating derivative and cumulative features of the PPG signal, our method captures physiologically meaningful dynamics that are critical for accurate ECG reconstruction. This multi-channel representation offers deeper insights into the temporal and morphological relationships between PPG and ECG signals.
    
    \item We demonstrate the efficacy of ViTs for ECG reconstruction from structured 2D PPG representations. The ViT effectively captures intra-beat morphology and inter-beat temporal dependencies via self-attention, outperforming state-of-the-art 1D sequence-based models in robustness and accuracy.

    \item We introduce new evaluation metrics: QRS area error, PR interval error, RT interval error, and RT amplitude difference, to assess ECG reconstruction quality beyond conventional metrics such as RMSE and HRV. These metrics provide a more nuanced and clinically interpretable assessment of waveform fidelity.
\end{enumerate}

\section{Related Work}
This section reviews recent advancements in methods for ECG reconstruction from PPG signals using 1D sequence-based approaches and introduces the Vision Transformer, which serves as the foundation of our proposed method.

\subsection{ECG Reconstruction Using 1D PPG Signals}
Significant progress has been made in ECG reconstruction from PPG signals through advanced deep learning methods. \cite{Sarkar2021} introduced CardioGAN, a model based on the Generative Adversarial Network (GAN) architecture \cite{goodfellow2014gan}. Inspired by CycleGAN \cite{zhuCycleGAN2017}, CardioGAN employs cycle consistency loss to train without requiring paired ECG-PPG data, showcasing the potential of GAN-based approaches for this task.

Building on diffusion models \cite{ho2020diffusion}, Shome et al. \cite{Shome2024} proposed the Region-Disentangled Diffusion Model (RDDM) for PPG-to-ECG translation. RDDM addresses a key limitation of traditional diffusion models, the indiscriminate application of noise across the entire signal, by introducing a region-specific noise process. This process selectively targets critical regions of interest (ROIs), such as the QRS complex in ECG signals, while preserving other parts of the waveform. By disentangling these regions, RDDM generates high-quality ECG signals from PPG inputs in ten diffusion steps \cite{Shome2024}.

\cite{Li2025underreview} introduced CLEP-GAN (Contrastive Learning for ECG Reconstruction from PPG Signals), incorporating contrastive learning, adversarial learning, and attention gating to facilitate precise, subject-independent ECG reconstruction. Although CLEP-GAN exhibits superior performance compared to previous methods, like other 1D convolution-based approaches, it encounters difficulties in accurately reconstructing smaller waveform features, such as the P-wave and T-wave. These features are crucial for detailed clinical assessment and diagnosis, like serum potassium and calcium estimation. To address these limitations, we propose a novel approach that incorporates the PPG signal's first-order difference, second-order difference, and AUC to enrich the input data, aiming to improve the reconstruction of small waveform features. 

\subsection{Vision Transformer}
The Vision Transformer (ViT) introduces a novel approach to computer vision by adapting the transformer architecture, originally developed for natural language processing, to image recognition tasks. Unlike traditional CNNs, which rely on localized convolutional operations to extract features, ViT employs self-attention mechanisms to model global context across the entire image.

In ViT, an input image with dimensions \(H \times W \times C\) (height, width, channels) is divided into a grid of non-overlapping patches. Each patch consists of \(P \times P\) pixels, where \(P\) is a fixed patch size (e.g., \(8 \times 8\)). The total number of patches, \(N\), is computed as $ N = \frac{H}{P} \times \frac{W}{P}$.Each patch is then flattened into a 1D vector of size \(P^2 \times C\) by concatenating its pixel values, effectively treating each patch as a token.

To encode spatial information, positional embeddings \(\mathbf{p}_i\) are added to each token embedding \(\mathbf{z}_0^i\), enabling the model to retain patch order and spatial relationships. The token embeddings are computed as:
\begin{align*}
    \mathbf{z}_0^i = \mathbf{E} \cdot \text{Flatten}(\text{Patch}_i) + \mathbf{p}_i, \quad i \in \{1, \dots, N\},
\end{align*}
% where \(\mathbf{E}\) is a learnable linear projection matrix, and \(\mathbf{p}_i\) represents the positional embedding of the \(i\)-th patch.
where \( \mathbf{E} \in \mathbb{R}^{D \times P^2C} \) is a learnable linear projection matrix, and \( \mathbf{p}_i \in \mathbb{R}^D \) is the positional embedding of the \( i \)-th patch. Here, \( P^2 \) is the patch size,  \( D \) is the embedding dimension, and \( C \) is the number of image channels. 

The position-augmented token embeddings are passed through a transformer encoder, which consists of multiple layers of multi-head self-attention and feed-forward networks. At the core of self-attention is the scaled dot-product attention mechanism, which computes the relationships between tokens as:
\begin{align*}
\text{Attention}(\mathbf{Q}, \mathbf{K}, \mathbf{V}) = \text{Softmax}\left(\frac{\mathbf{Q} \mathbf{K}^T}{\sqrt{d_k}}\right) \mathbf{V},
\end{align*}
% where \(\mathbf{Q}\), \(\mathbf{K}\), and \(\mathbf{V}\) are the query, key, and value matrices derived from the token embeddings. \( d_k \) is the dimension of \(\mathbf{K}\). 

where \( \mathbf{Q} = \mathbf{x} \mathbf{W}^Q \), \( \mathbf{K} = \mathbf{x} \mathbf{W}^K \), and \( \mathbf{V} = \mathbf{x} \mathbf{W}^V \) are the query, key, and value matrices, respectively, computed from the input sequence \( \mathbf{x} \in \mathbb{R}^{B \times T \times D} \). The projection matrices are \( \mathbf{W}^Q, \mathbf{W}^K, \mathbf{W}^V \in \mathbb{R}^{D \times D} \), and the resulting \( \mathbf{Q}, \mathbf{K}, \mathbf{V} \) all lie in \( \mathbb{R}^{B \times T \times D} \). The scalar \( d_k \) denotes the key dimensionality, where \( B \) is the batch size, \( T \) is the number of tokens, \( D \) is the embedding dimension.

This mechanism enables the model to focus on relevant patches while considering their global relationships. Self-attention is extended to multi-head self-attention, where multiple attention heads operate in parallel:
\begin{align*}
\text{MultiHead}(\mathbf{Q}, \mathbf{K}, \mathbf{V}) = \text{Concat}(\text{head}_1, \dots, \text{head}_h) \mathbf{W}^O,
\end{align*}
with each attention head defined as $\text{head}_i = \text{Attention}(\mathbf{Q} \mathbf{W}_i^Q, \mathbf{K} \mathbf{W}_i^K, \mathbf{V} \mathbf{W}_i^V), \quad i = 1, \dots, h,$
% \begin{align*}
% \text{head}_i = \text{Attention}(\mathbf{Q} \mathbf{W}_i^Q, \mathbf{K} \mathbf{W}_i^K, \mathbf{V} \mathbf{W}_i^V), \quad i = 1, \dots, h,
% \end{align*}
where the projection matrices are \( \mathbf{W}_i^Q, \mathbf{W}_i^K, \mathbf{W}_i^V \in \mathbb{R}^{D \times d_k} \). Here, \( d_k = d_v = D / h \), and \( h \) is the number of attention heads. After computing the individual heads, the outputs are concatenated along the last dimension, yielding a tensor of shape \( \mathbb{R}^{B \times T \times D} \). This concatenated output is then projected back to the original embedding dimension using an output projection matrix \( \mathbf{W}^O \in \mathbb{R}^{D \times D} \). 

To stabilize training and facilitate optimization, residual connections and layer normalization are applied following the self-attention operation. The output of the multi-head attention layer is given by $\mathbf{z}_{\text{att}} = \text{MultiHead}(\mathbf{Q}, \mathbf{K}, \mathbf{V}) + \mathbf{z}_{\text{in}},$
% \begin{align*}
% \mathbf{z}_{\text{att}} = \text{MultiHead}(\mathbf{Q}, \mathbf{K}, \mathbf{V}) + \mathbf{z}_{\text{in}},
% \end{align*}
where \( \mathbf{z}_{\text{in}} \) denotes the input token embeddings to the transformer layer, and \( \mathbf{z}_{\text{att}} \) is the output after applying multi-head self-attention.

To further refine the representation, the attention output is passed through a feed-forward network (FFN): $\mathbf{z}_{\text{out}} = \text{FFN}(\mathbf{z}_{\text{att}}),$
% \begin{align*}
% \mathbf{z}_{\text{out}} = \text{FFN}(\mathbf{z}_{\text{att}}),
% \end{align*}
where the FFN consists of two fully connected layers with a non-linear activation function, defined as $ \text{FFN}(\mathbf{z}_{\text{att}}) = \text{ReLU}(\mathbf{z}_{\text{att}} \mathbf{W}_1 + \mathbf{b}_1)\mathbf{W}_2 + \mathbf{b}_2.$
% \begin{align*}
%     \text{FFN}(\mathbf{z}_{\text{att}}) = \text{ReLU}(\mathbf{z}_{\text{att}} \mathbf{W}_1 + \mathbf{b}_1)\mathbf{W}_2 + \mathbf{b}_2,
% \end{align*}
Here, \( \mathbf{z}_{\text{att}} \in \mathbb{R}^{B \times T \times D} \) denotes the output of the attention sub-layer. The weight matrix of the first linear layer is \( \mathbf{W}_1 \in \mathbb{R}^{D \times H} \), and its corresponding bias is \( \mathbf{b}_1 \in \mathbb{R}^{H} \), where \( H \) is the hidden dimension of the feed-forward network. The second linear layer uses weight matrix \( \mathbf{W}_2 \in \mathbb{R}^{H \times D} \) and bias \( \mathbf{b}_2 \in \mathbb{R}^{D} \).

By treating image patches as tokens, ViT processes images similarly to how transformers process sequences of words, allowing it to capture both local and global dependencies. Unlike CNNs, which incorporate strong inductive biases such as locality and translation equivariance, ViT learns such features purely from data. Its self-attention mechanism is particularly effective at modeling long-range dependencies between distant patches, allowing for a more holistic understanding of global image structure, an aspect that CNNs may struggle to represent. This capability makes ViT especially well-suited for ECG reconstruction, where both fine-grained waveform morphology and global rhythm patterns must be accurately captured.

\section{Dataset} \label{sec:dataset}
We evaluated the efficacy of our method using two publicly available real-world datasets. The BIDMC PPG and Respiration Dataset contains 53 paired PPG and ECG records from 45 patients monitored at Beth Israel Deaconess Medical Center \cite{Pimentel2017Towards, Goldberger2000PhysioBank}. Each record has a duration of 8 minutes, sampled at 125 Hz. The dataset includes data from 20 male patients aged between 19 and over 90 years, with a mean age of 66 years (SD: 17). For further validation, we utilized the CapnoBase TBME RR Benchmark Dataset \cite{karlen2013capnobase}, comprising 42 eight-minute PPG and ECG records sampled at 300 Hz. These records were sourced from 29 pediatric surgeries and 13 adult surgeries, each corresponding to a unique individual.

The CapnoBase and BIDMC datasets are widely used in biomedical signal processing research but differ significantly in terms of their patient populations, data collection settings, and signal characteristics. The primary purpose of the CapnoBase dataset is to serve as a benchmark for respiratory rate estimation algorithms, emphasizing clean, well-annotated signals with minimal noise \cite{charlton2020capnobase}. In contrast, the BIDMC dataset is derived exclusively from critically ill patients in intensive care unit (ICU) or sleep study settings, focusing on real-world clinical conditions. It features PPG, ECG, and respiratory signals, often exhibiting significant noise, irregularities, and pathological patterns \cite{Goldberger2000PhysioBank}. While CapnoBase offers clean signals ideal for benchmarking, BIDMC provides a more realistic dataset for testing algorithms in noisy and variable environments typical of ICUs, making these datasets complementary for algorithm development and validation \cite{charlton2020capnobase, Goldberger2000PhysioBank}.
 
\subsection{Data processing} \label{sec:data_processing}
To reduce noise in the real-world datasets, we applied bandpass filters to both ECG and PPG signals. The ECG signals were filtered within a frequency range of 0.4 Hz to 45 Hz to effectively capture essential cardiac events such as the P-wave, QRS complex, and T-wave, which primarily occur between 0.5 Hz and 100 Hz. This range ensures a balance between excluding higher frequency noise and retaining critical signal components necessary for accurate analysis \cite{Li2025underreview}.

Similarly, for PPG signals, we used a bandpass filter ranging from 0.3 Hz to 8 Hz. This frequency range is optimal for preserving vital physiological features like heart rate and respiratory rate, while efficiently reducing extraneous noise. This approach aligns with established practices in the field, as documented in previous studies \cite{Shelley2007, Allen2007} \cite{Li2025underreview}.

\section{Method} \label{sec:method}
\begin{figure}[!ht] 
\centering
\begin{subfigure}[b]{0.7\columnwidth}
  \centering
    \includegraphics[width=\linewidth]{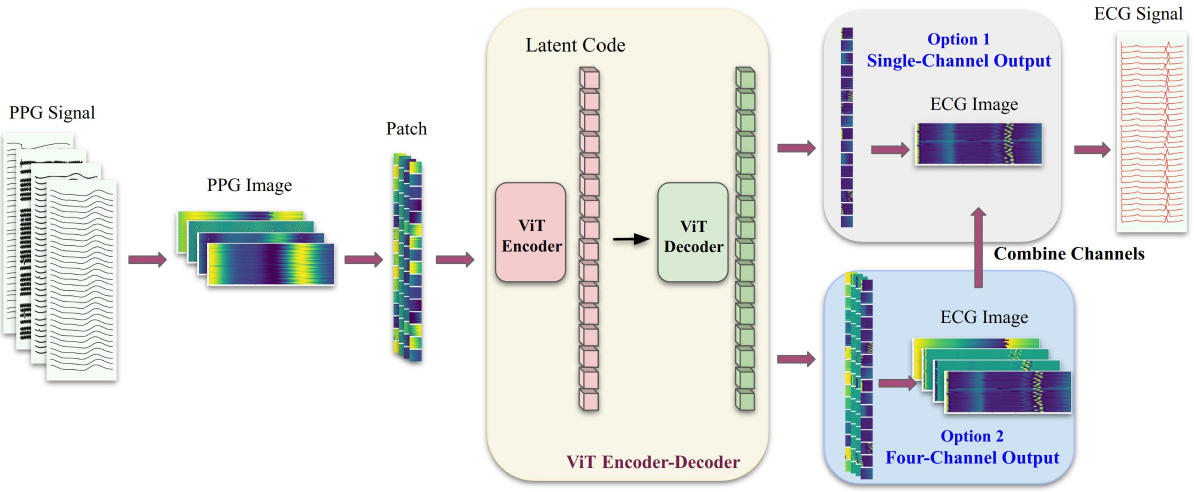}
  \caption{Main architecture.}
  \label{fig:main_architecture} 
\end{subfigure}
\hfill
\begin{subfigure}[b]{0.7\columnwidth}
  \centering
    \includegraphics[width=\linewidth]{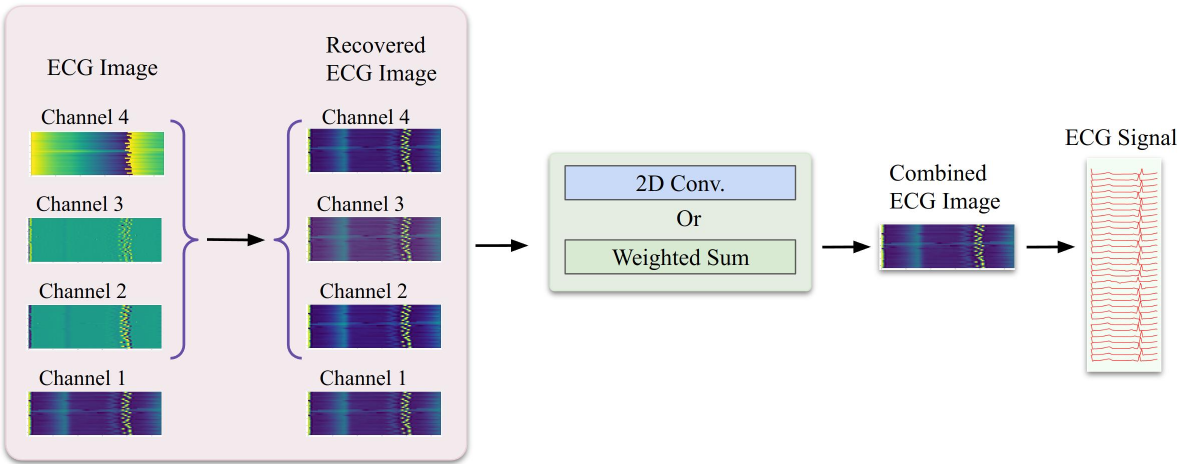}
  \caption{Combine four individual images into a 1D ECG signal.}
  \label{fig:combination} 
\end{subfigure}
\caption{Architecture of the proposed method. (a) presents the main framework, while (b) illustrates the process of merging the four output channels from the ViT decoder into the final ECG image. This process involves reconstructing three intermediate ECG images by reversing the first-order difference image (channel 2), the second-order difference image (channel 3), and the AUC image (channel 4). These intermediate signals are then combined with the reconstructed ECG image (channel 1) using one of two methods: a 2D convolution layer or a weighted sum. Both the encoder and decoder are implemented using(ViT networks.}
\label{fig:architecture}.
\vspace{-0.1in}
\end{figure}

Figure \ref{fig:architecture} illustrates the architecture of our proposed methods. Specifically, Figure \ref{fig:main_architecture} presents the overall framework, which employs an encoder-decoder framework. Both the encoder and decoder are implemented using ViT networks. The input PPG signal is first combined with its first-order difference, second-order difference, and AUC to form a four-channel time series. This multi-channel time series is then transformed into a 2D format, creating a four-channel 2D input image. The encoder-decoder network processes this input image to generate either a single-channel or four-channel ECG image.

When the ViT decoder outputs a four-channel ECG image, an additional transformation is required to obtain the final 1D ECG signal. To achieve this, we propose two approaches: (1) applying a 2D convolution layer to combine the four channels, and (2) computing a weighted sum of the four channels. These combination methods are illustrated in Figure \ref{fig:combination}. In the weighted sum approach, the weights for each channel are learnable parameters that are optimized during training.

\subsubsection{2D Signal Image Data}
In our proposed method, we first compute the first-order difference, second-order difference, and cumulative AUC for the PPG ($X$) and ECG ($Y$) signal. Each resulting 1D time series is then reshaped into a 2D image, and the four resulting images are combined to form a single four-channel image.

\paragraph{First-Order Difference}  
The first-order difference of the PPG signal is calculated as:  
\begin{align}
    \Delta y[n] = \frac{y[n+1] - y[n]}{\Delta t},
\end{align}  
where \(\Delta t\) represents the time interval between consecutive time steps, and \(n\) is the time step index. To maintain the original sequence length, the last element is padded with the value from the previous time step, creating the second channel of the time series.

\paragraph{Second-Order Difference}  
The second-order difference of the signal is computed as:  
\begin{align}
    \Delta^2 y[n] = \frac{\Delta y[n+1] - \Delta y[n]}{\Delta t}.
\end{align}  
Similar to the first-order difference, the sequence is padded at the end to maintain the original length, forming the third channel of the time series.

\paragraph{Cumulative AUC}  
The cumulative AUC of the signal is calculated using the trapezoidal rule:  
\begin{align}
    y_{\text{AUC}}[n] = \sum_{m=0}^{n-1} \frac{y[m] + y[m+1]}{2} \Delta t.
\end{align}  
To represent the starting point of the integration, a 0 is prepended to the sequence. This forms the fourth channel time sequence.

\paragraph{Transforming to Four-Channel Signal Image}  

To construct the ECG image, we transform each channel \(\left(y[n], \Delta y[n], \Delta^2 y[n], y_{\text{AUC}}[n]\right)\) into a 2D representation, where each row corresponds to a single cardiac beat. Since R peak detection is generally more reliable than identifying the onsets or offsets of P and T waves, we use the RR interval to define each beat cycle. Due to the variability in beat durations, we first determine the longest beat in the dataset and pad all shorter beats with values from subsequent beats. Consequently, each segment starts at an R peak and spans a window equal to the longest beat length in the dataset. 

\begin{figure}[htbp!]
  \centering
  \begin{subfigure}[b]{0.3\columnwidth}
    \centering
    \includegraphics[width=0.4\linewidth]{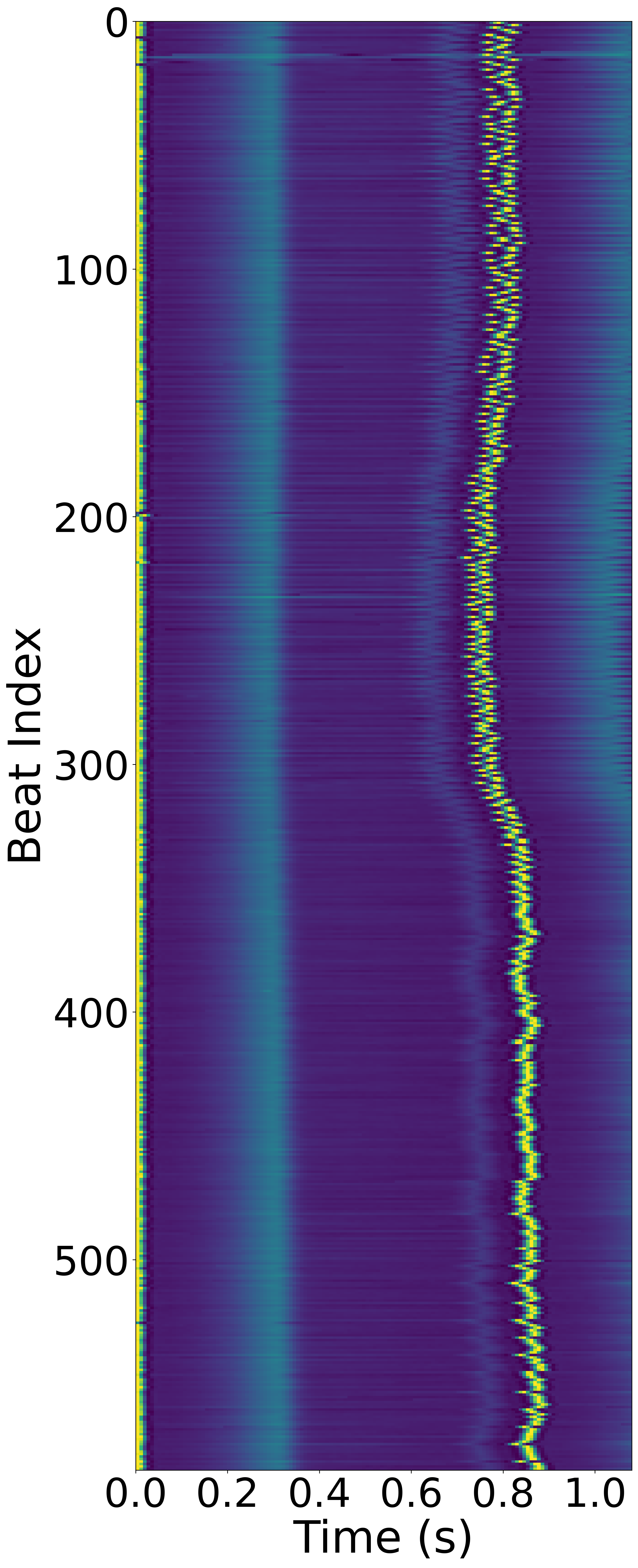}
    \caption{ECG}
    \label{fig:ecg_beat_0332}
  \end{subfigure}
  % \hspace{0.01\columnwidth} 
  \begin{subfigure}[b]{0.3\columnwidth}
    \centering
    \includegraphics[width=0.4\linewidth]{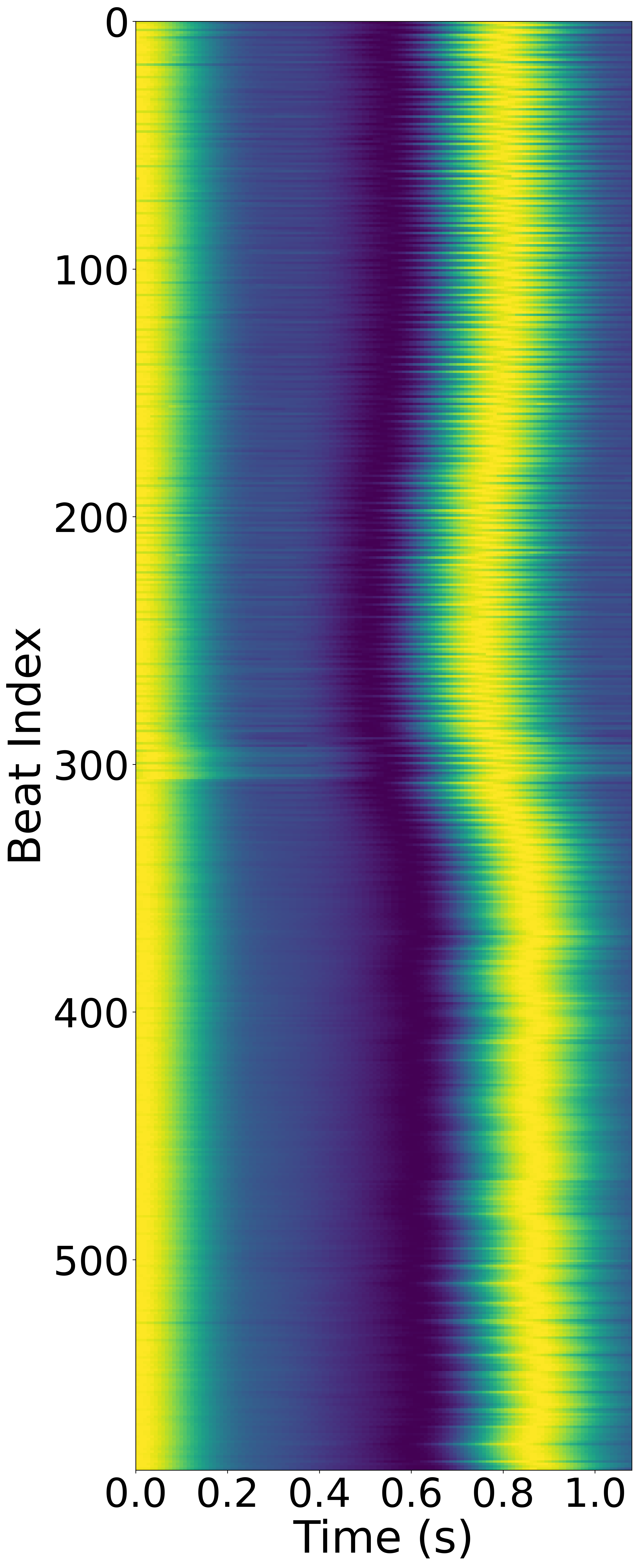}
    \caption{PPG}
    \label{fig:ppg_beat_0332}
  \end{subfigure}
  \caption{Examples of padded ECG and PPG beats from record 0332 in the CapnoBase dataset.}
  \label{fig:padded_beats}
  \vspace{-0.2in}
\end{figure}

Figure~\ref{fig:padded_beats} illustrates padded ECG and PPG beats from record 0332 in the CapnoBase dataset. In the ECG beats (Figure~\ref{fig:ecg_beat_0332}), the light-colored regions correspond to the R peaks, P waves, and T waves. In the PPG beats (Figure \ref{fig:ppg_beat_0332}), the high-intensity regions primarily reflect the systolic phase, with the light blue areas marking the dicrotic notch. To construct the final 2D image, we stack 16 consecutive padded beats. While increasing the number of beats per image could enhance prediction robustness, excessively large stacks reduce the effective training dataset size, potentially limiting generalization. This process is applied individually to each channel to generate its corresponding 2D signal image.

The 2D ECG signal images for all four channels are then stacked along the channel dimension to form the final four-channel image:
\begin{align}
Y_{I} = 
\begin{bmatrix}
Y_{1C} \\
Y_{\Delta x} \\
Y_{\Delta^2 x} \\
Y_{{\text{AUC}}}
\end{bmatrix},
\end{align}  
where, \(Y_{1C}\) is 2D representation of the original signal. \(Y_{\Delta x}\) represents 2D representation of the signal's first-order difference. \(Y_{\Delta^2 x}\) denotes 2D representation of the signal's second-order difference. \(Y_{{\text{AUC}}}\) represents 2D representation of the signal's cumulative AUC. The same processing steps used for the ECG image ($Y_I$) are applied to the PPG image ($X_I$).

\subsection{Objective} \label{sec:objective}
Our approach employs a ViT encoder-decoder architecture optimized using reconstruction losses. We utilize the mean squared error (MSE) as the primary reconstruction loss. Additionally, to further improve reconstruction quality, we incorporate the QRS complex-enhanced loss introduced by \cite{Chiu2020}.
\subsubsection{Reconstruction Loss}
The reconstruction loss measures the accuracy of the generated ECG image (\(\hat{Y}_{1C}\)) by comparing it to the ground-truth ECG image (\(Y_{1C}\)). It is defined as the MSE between the predicted and ground-truth single-channel ECG images:
\begin{align}
\mathcal{L}_{\text{R}_{1C}} = \frac{1}{P} \sum_{i=1}^{P} \left( \hat{Y}_{1C}[i] - Y_{1C}[i] \right)^2,
\end{align}
where \(P\) represents the total number of pixels in the single-channel ECG image.

\subsubsection{QRS Complex-Enhanced Loss}  
To improve reconstruction accuracy, we incorporate a weighted loss function that prioritizes the QRS region, as introduced by \cite{Chiu2020}. This function applies a Gaussian weighting centered around R-peak locations, ensuring that reconstruction errors within the QRS complex are penalized more heavily. The Gaussian weight is controlled by a spread parameter \(\sigma\) and an intensity factor \(\beta\), and the loss is defined as:
\begin{align}
\mathcal{L}_{\text{QRS}} = \sum_{l=1}^{L} \sum_{t=1}^{T} \left| y_{l,t} - \hat{y}_{l,t} \right| \left( 1 + \beta \sum_{k=1}^{K_l} e^{-\frac{(t - c_{l,k})^2}{2\sigma^2}} \right),
\end{align}
where \( c_{l,k} \) denotes the location of the \(k\)-th R-peak in the \(l\)-th ECG segment, and \( K_l \) is the total number of R-peaks in that segment. \( L \) represents the number of ECG segments in the batch, and \( T \) is the number of time steps in each segment. Since this loss function operates in the temporal domain, the generated ECG image must first be converted back into a 1D signal before computing the QRS-enhanced loss.

\subsubsection{Total Loss}
The total loss function used to optimize the ViT encoder-decoder and the channel combination layer is defined as the sum of the reconstruction loss and the QRS complex-enhanced loss:
\begin{align}
\mathcal{L}_{\text{total}} = \mathcal{L}_{\text{R}_{1C}} + \mathcal{L}_{\text{QRS}}.
\end{align}

\section{Experiments} \label{sec:experiments}
\subsection{Experimental Setup}
Given a sample rate of 125 Hz, the time step was set to 1 for computing the first-order difference, second-order difference, and AUC. The signal images had a resolution of \(16 \times 128\), where 16 represents the number of beats and 128 corresponds to the beat length. In our experiments, the patch size for the ViT model was set to 8. The ViT encoder consisted of 12 transformer layers, each structured as a block with three attention heads, while the ViT decoder comprised 8 layers, also featuring three attention heads per layer. Training was conducted with a batch size of 64 using the AdamW optimizer (\(\text{lr} = 1.5 \times 10^{-4}\), weight decay = 0.05) for 2000 epochs. All experiments were performed on an NVIDIA Quadro RTX 6000 GPU with 24 GB of memory.

\subsection{Evaluation Metrics}
\paragraph{Root Mean Square Error (RMSE)}
RMSE quantifies the difference between the predicted and actual signals. It is formally defined as:
\begin{align}
\text{RMSE} = \sqrt{\frac{\sum_{i=1}^N \left(Y[i] - \hat{Y}[i]\right)^2}{N}},
\end{align}
where \(Y\) represents the ground-truth signal, \(\hat{Y}\) is the reconstructed signal, and \(N\) denotes the length of the signal or the number of data points in the signal.

\paragraph{Percentage Root Mean Square Difference (PRD)}  
PRD measures the relative disparity between the reconstructed signal and the ground-truth signal as a percentage. It is formally defined as:
\begin{align}
\text{PRD} = \sqrt{\frac{\sum_{i=1}^N \left(Y[i] - \hat{Y}[i]\right)^2}{\sum_{i=1}^N \left(Y[i]\right)^2}} \times 100,
\end{align}
here, \(N\) denotes the length of the signal.

\paragraph{Heart Rate Variability (HRV)}
HRV quantifies the variability in time intervals between consecutive heartbeats, reflecting autonomic nervous system activity and cardiovascular health. In our experiments, HRV was assessed using statistical metrics such as the mean and standard deviation (STD) of RR intervals, defined as the time between successive R peaks in the ECG waveform. Importantly, all RR intervals were included in the analysis without excluding ectopic or abnormal beats, departing from standard HRV methodologies that rely on normal-to-normal intervals to exclude arrhythmic variability.

In addition to RR intervals, this study also analyzes HRV using PP intervals (time between consecutive P waves) and TT intervals (time between consecutive T waves). This comprehensive approach provides additional insights into the variability of different segments of the cardiac cycle.

\paragraph{QRS Area Error}  
The QRS area for a single QRS complex is defined as:  
% \begin{align}
% A_{\text{QRS}} = \int_{\text{QRS}_B}^{\text{QRS}_E} Y(t) \, dt,
% \end{align}
\begin{align}
A_{\text{QRS}} = \sum_{i = t_s}^{t_e} Y[i]
\end{align}

where \(Y[i]\) represents the ECG signal, and \(t_s\) and \(t_e\) denote the time indices marking the beginning and end of the QRS complex, respectively.

For multiple QRS complexes, the mean QRS area of the ECG signal \(Y_{A_{QRS}}\) is computed as:
% and the reconstructed signal (\(\text{RA}\)) are computed as:  
\begin{align}
Y_{A_{\text{QRS}}} = \frac{1}{B} \sum_{b=1}^{B} A_{\text{QRS}}[b],
\end{align}
where \( B \) represents the total number of ECG beats (or QRS complexes). The relative QRS area error is then defined as:  
\begin{align}
\text{RE}_{\text{QRS}} = \frac{\left|Y_{A_{\text{QRS}}} - \hat{Y}_{A_{\text{QRS}}} \right|}{Y_{A_{\text{QRS}}}}.
\end{align}

\paragraph{PR Interval Error}  
The PR interval represents the time between the onset of atrial depolarization (P wave) and the onset of ventricular depolarization (R wave). In this study, to mitigate the impact of noise interference, which can hinder the accurate detection of small wave onsets, we use the interval between the P peak and R peak as a practical approximation of the true PR interval. Peaks are generally easier to identify and locate with greater precision in ECG signals, making this approach more reliable for analysis. The PR interval, expressed in milliseconds, is calculated as:
\begin{align}
Y_{\text{PR}} = \frac{1}{B} \sum_{b=1}^{B} \left( \frac{R[b] - P[b]}{f_s} \times 1000 \right),
\end{align}
where \( f_s \) is the ECG sampling rate (in Hz). \( P[b] \) and \( R[b] \) represent the sample indices of the \(b\)-th P peak and R peak, respectively. To evaluate the reconstruction accuracy, the PR relative error is defined as:
\begin{align}
\text{RE}_{\text{PR}} = \frac{\left| Y_{\text{PR}} - \hat{Y}_{\text{PR}} \right|}{Y_{\text{PR}}}.
\end{align}

\paragraph{RT Interval Error}  
The interval from the R peak to the T peak, referred to as the RT interval in this study, encompasses both the ST segment and the T-wave, reflecting the entire period of ventricular repolarization. Typically, the ST segment, running from the end of the S-wave to the start of the T-wave, is critical for evaluating ventricular repolarization. However, in cases where the S-wave is indistinct or difficult to identify in certain beats, accurately delineating the ST segment becomes challenging. Therefore, in our analysis, we approximate the assessment of ventricular repolarization by measuring from the R peak to the T peak. The RT interval, expressed in milliseconds, is calculated as:
\begin{align}
Y_{\text{RT}} = \frac{1}{B}\sum_{b=1}^{B}\left(\frac{T[b] - R[b]}{f_s} \times 1000\right),
\end{align}
here \( R[b] \) and \( T[b] \) represent the sample indices of the \(b\)-th R peak and T peak, respectively. The RT relative error is calculated as:
\begin{align}
\text{RE}_{\text{RT}} = \frac{\left| Y_{\text{RT}} - \hat{Y}_{\text{RT}} \right|}{Y_{\text{RT}}}.
\end{align}

\paragraph{RT Amplitude Difference}  
The RT Amplitude Difference measures the difference in amplitude between the R peak, which represents ventricular depolarization, and the T peak, which represents ventricular repolarization. This metric is used to evaluate the relationship between the depolarization and repolarization phases of the cardiac cycle. The RT Amplitude Difference (\(\text{AD}_{\text{RT}}\)) is calculated as:
\begin{align}
\text{AD}_{\text{RT}} = \frac{1}{B} \sum_{b=1}^{B} \left| A_{\text{R}}[b] - A_{\text{T}}[b] \right|,
\end{align}
\noindent
where \( A_{\text{R}}[b] \) is the amplitude of the R peak, and \( A_{\text{T}}[b] \) is the amplitude of the T wave for the \(b\)-th beat. The relative error for RT amplitude difference is defined as:
\begin{align}
\text{RE}_{\text{AD}} = \frac{\left| \text{AD}_{\text{RT}} - \widehat{\text{AD}}_{\text{RT}} \right|}{\text{AD}_{\text{RT}}},
\end{align}
where \( \text{AD}_{\text{RT}} \) and \( \widehat{\text{AD}}_{\text{RT}} \) denote the RT Amplitude Difference for the original and reconstructed signals, respectively. This metric is useful for identifying myocardial injury or ischemia, which can alter the amplitude and morphology of both the R and T waves \cite{surawicz2009}. For example, ischemia may suppress the R wave or elevate the T wave, while ventricular hypertrophy can amplify the R wave amplitude and distort repolarization, leading to deviations in the RT amplitude difference \cite{Bornstein2025}.

\subsection{Experimental Results} \label{sec:results}
We evaluated 30 record pairs from each dataset: BIDMC and CapnoBase, using a leave-one-out validation strategy, where one record served as the test set and the remaining were used for training. Because our evaluation metrics, including QRS area error, PR interval error, RT interval error, and RT amplitude difference, depend on peak-detection algorithms, and small waveforms (such as P waves) are particularly susceptible to noise, we selected 30 relatively low-noise recordings from each dataset to ensure reliable measurement accuracy. As a baseline, we employed the state-of-the-art 1D convolutional model CLEP-GAN \cite{Li2025underreview}, chosen for its superior performance compared to other advanced models. Table \ref{tab:leave_one_out} summarizes the average performance on both datasets. All test results for our proposed method were based on the single-channel output structure shown in Figure \ref{fig:architecture}.

\begin{figure}[htbp!] 
\centering
\begin{subfigure}{0.45\textwidth}
  \centering
    \includegraphics[width=\linewidth]{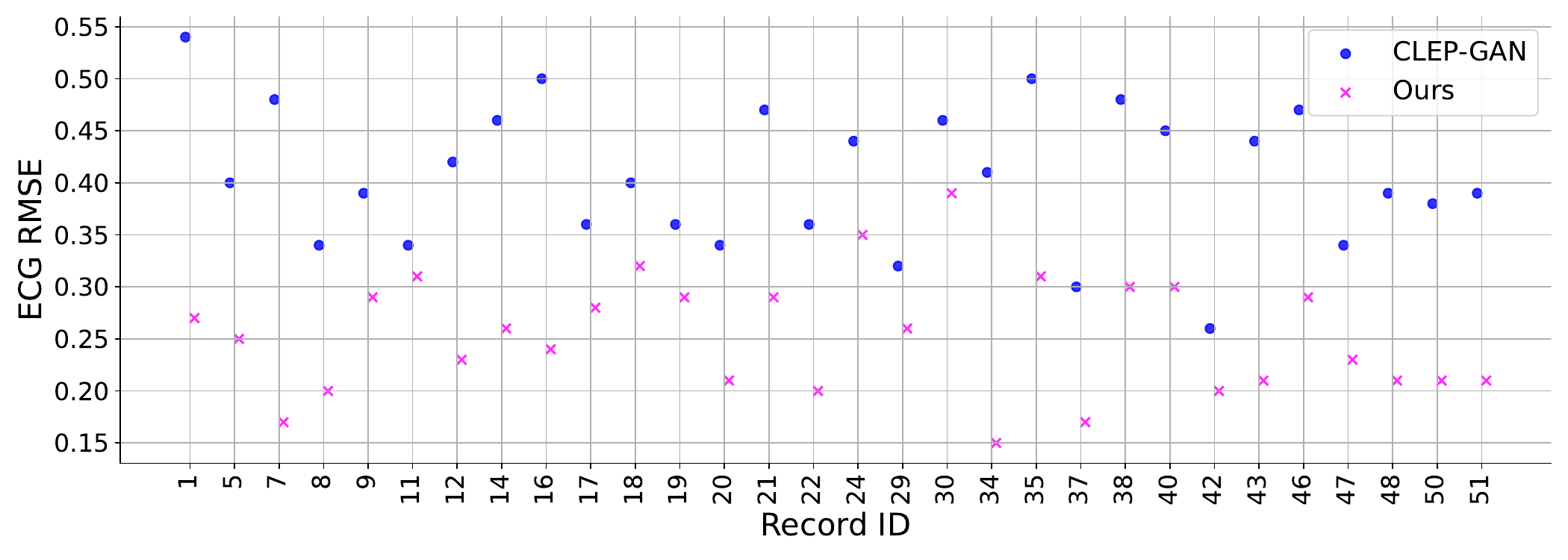}
  \caption{RMSE.}
  \label{fig:plot_bidmc_rmse} 
\end{subfigure}
\begin{subfigure}{0.45\textwidth}
  \centering
    \includegraphics[width=\linewidth]{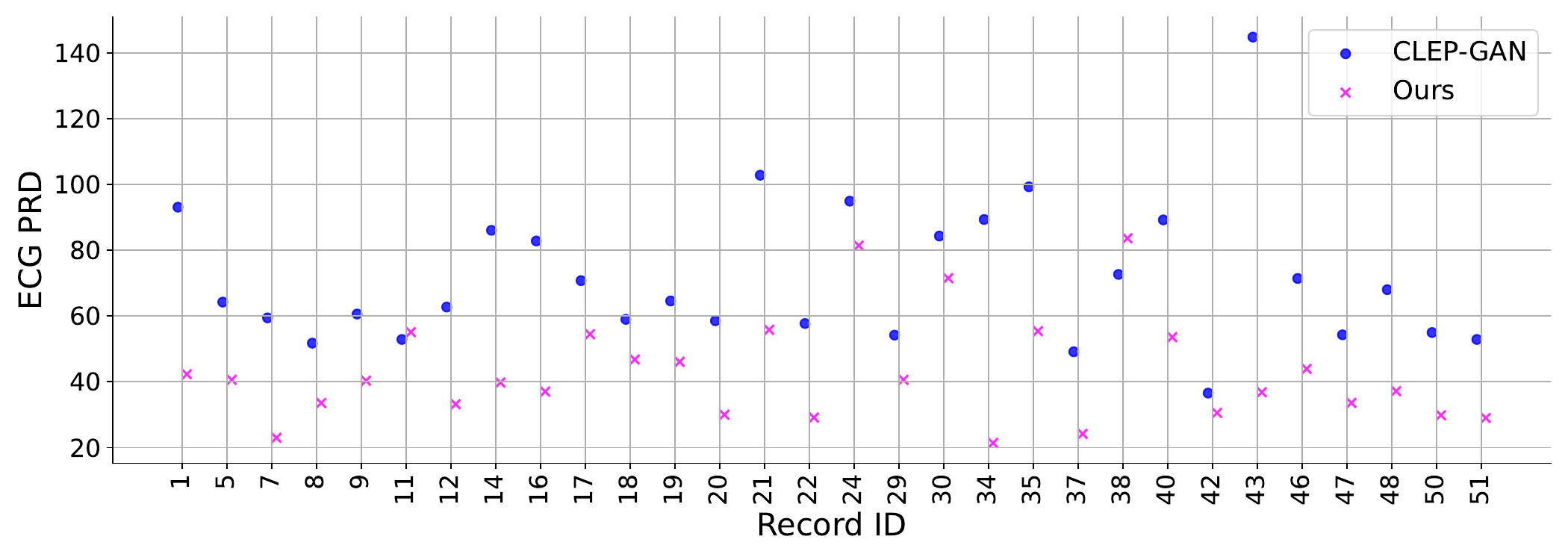}
  \caption{PRD.}
  \label{fig:plot_bidmc_prd} 
\end{subfigure}
\caption{Scatter plots of RMSE and PRD results for CLEP-GAN and our method on the BIDMC dataset.}
\label{fig:plot_bidmc_rmse_prd}.
\vspace{-0.1in}
\end{figure}

\begin{figure}[htbp!] 
\vspace{-0.2in}
\centering
\begin{subfigure}{0.45\textwidth}
  \centering
    \includegraphics[width=\linewidth]{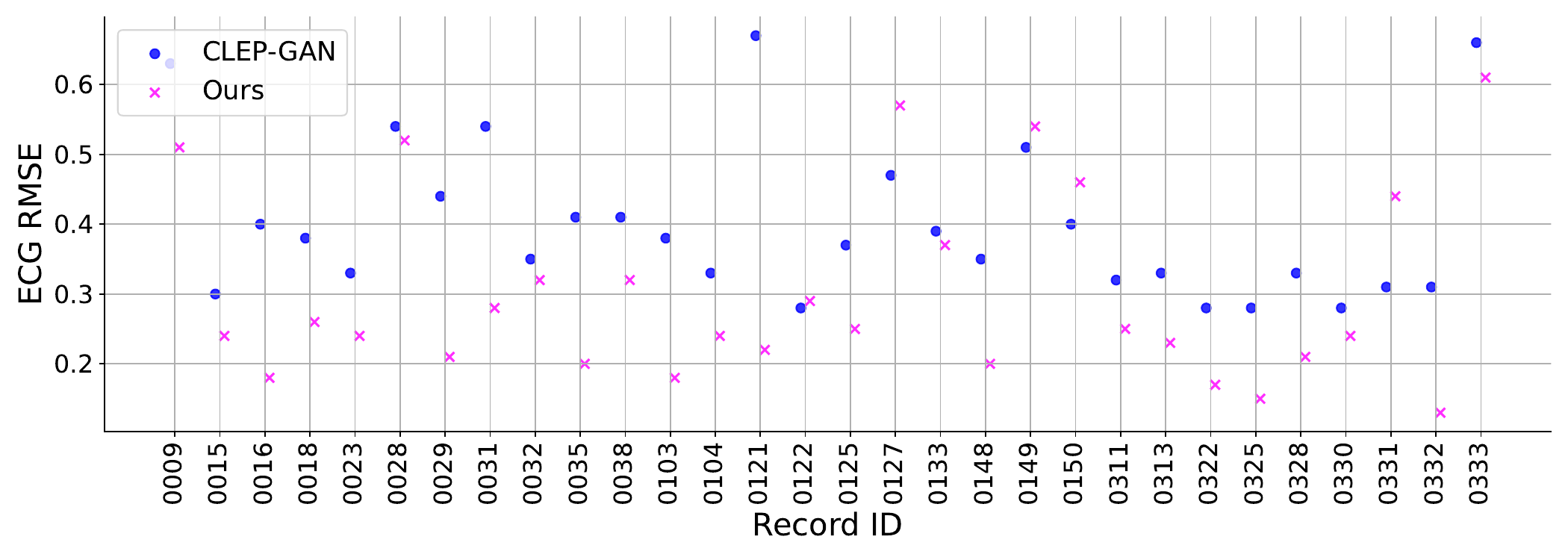}
  \caption{RMSE.}
  \label{fig:plot_capno_rmse} 
\end{subfigure}
\begin{subfigure}{0.45\textwidth}
  \centering
    \includegraphics[width=\linewidth]{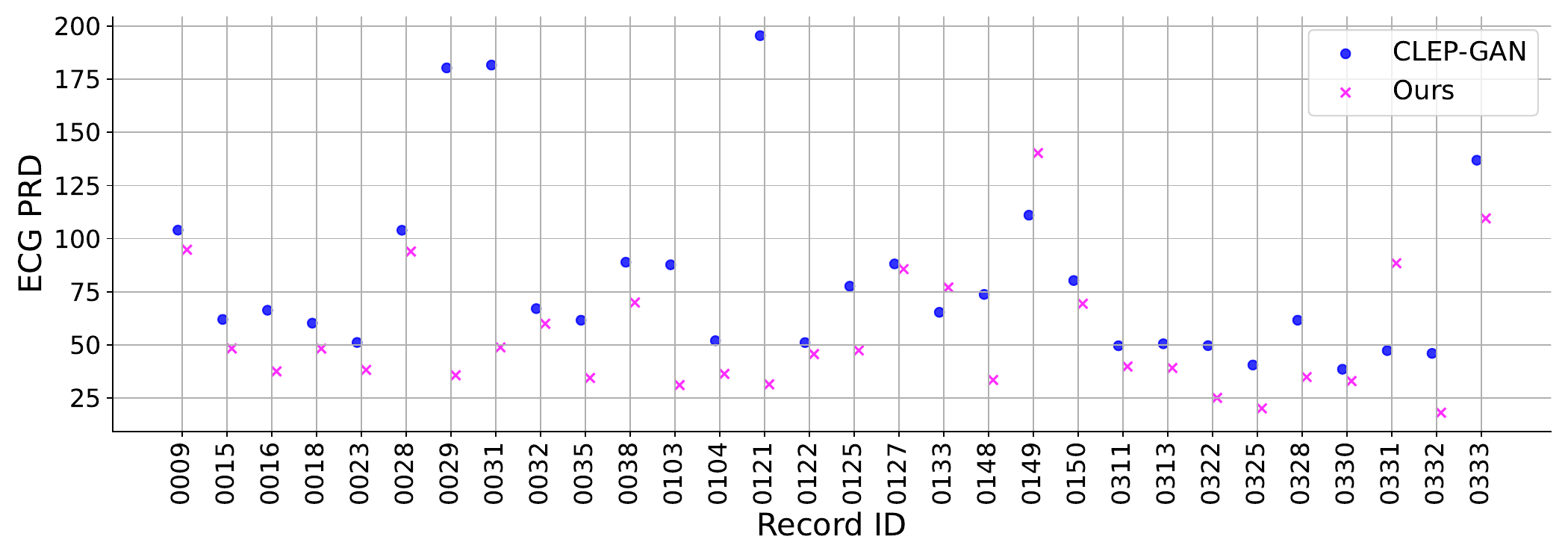}
  \caption{PRD.}
  \label{fig:plot_capno_prd} 
\end{subfigure}
\caption{Scatter plots of RMSE and PRD results for CLEP-GAN and our method on the CapnoBase dataset.}
\label{fig:plot_capno_rmse_prd}.
\vspace{-0.1in}
\end{figure}

The results in Table \ref{tab:leave_one_out} demonstrate that our method achieves significantly lower PRD values compared to CLEP-GAN, with reductions of approximately 29\% and 18\% on the BIDMC and CapnoBase datasets, respectively. The RMSE is also notably reduced, by approximately 15\% and 10\% on the two datasets. In addition, our method consistently outperforms CLEP-GAN across all other evaluation metrics for both datasets, except for a slightly higher relative error in QRS area (\(\text{RE}_{\text{QRS}}\)) on the CapnoBase dataset. The standard deviation values (shown in parentheses) further indicate that our method exhibits greater stability than CLEP-GAN. Figures \ref{fig:plot_bidmc_rmse_prd} and \ref{fig:plot_capno_rmse_prd}  present scatter plots of RMSE and PRD for individual records on the BIDMC and CapnoBase datasets, respectively. These plots reveal a consistent trend in which our method outperforms CLEP-GAN across the majority of cases. 
% Additional results for other evaluation metrics are provided in Appendix \ref{sec:other_scatter_plots}.
\begin{table}[htbp!]
\centering
\footnotesize
%\begin{sidewaystable} % Rotate the entire table to landscape
\caption{Comparison of our method and CLEP-GAN on the BIDMC and CapnoBase datasets using leave-one-out validation. Values are reported as mean ($\pm$ standard deviation) across 30 records from each dataset. Evaluation metrics include the HRV for the R wave, T wave, and P wave intervals (R-HRV, T-HRV, and P-HRV, respectively), as well as the relative errors for QRS area (\(\text{RE}_{\text{QRS}}\)), PR interval (\(\text{RE}_{\text{PR}}\)), RT interval (\(\text{RE}_{\text{RT}}\)), and RT amplitude difference (\(\text{RE}_{\text{AD}}\)). All HRV values are reported in milliseconds. Lower values indicate better performance for all metrics.}
\label{tab:leave_one_out}
\begin{tabular}{l|l|l|l|l}
\hline
        & \multicolumn{2}{c|}{\textbf{BIDMC}} & \multicolumn{2}{c}{\textbf{CapnoBase}} \\ 
\textbf{Metric} & CLEP-GAN & Our & CLEP-GAN & Our \\
 \hline
PRD  $\downarrow$                       & 71.43 ($\pm 21.54$) & \textbf{42.65} ($\pm 15.48$) & 81.04 ($\pm 41.59$) & \textbf{53.89} ($\pm 28.70$) \\ \hline
RMSE  $\downarrow$                      & 0.406 ($\pm 0.066$) & \textbf{0.253} ($\pm 0.056$) & 0.399 ($\pm 0.111$) & \textbf{0.301} ($\pm 0.134$)\\  \hline
R-HRV $\downarrow$                      & 0.041 ($\pm 0.086$) & \textbf{0.01} ($\pm 0.029$) & 0.018 ($\pm 0.045$) & \textbf{0.008} ($\pm 0.022$) \\  \hline
T-HRV  $\downarrow$                     & 0.041 ($\pm 0.086$) & \textbf{0.01} ($\pm 0.029$) & 0.018 ($\pm 0.045$) & \textbf{0.008} ($\pm 0.022$) \\  \hline
P-HRV $\downarrow$                      & 0.041 ($\pm 0.086$) & \textbf{0.01} ($\pm 0.029$) & 0.019 ($\pm 0.045$) & \textbf{0.008} ($\pm 0.022$) \\  \hline
\(\text{RE}_{\text{QRS}}\) $\downarrow$ & 0.208 ($\pm 0.153$) & \textbf{0.191} ($\pm 0.149$) & \textbf{0.229} ($\pm 0.153$) & 0.260 ($\pm 0.161$) \\ \hline
\(\text{RE}_{\text{PR}}\) $\downarrow$  & 0.267 ($\pm 0.223$) & \textbf{0.214} ($\pm 0.184$) & 0.175 ($\pm 0.146$) & \textbf{0.128} ($\pm 0.1$) \\  \hline
\(\text{RE}_{\text{RT}}\) $\downarrow$  & 0.244 ($\pm 0.185$) & \textbf{0.19} ($\pm 0.136$) & 0.151 ($\pm 0.23$) & \textbf{0.146} ($\pm 0.219$) \\  \hline
\(\text{RE}_{\text{AD}}\) $\downarrow$  & 0.215 ($\pm 0.188$) & \textbf{0.155} ($\pm 0.207$) & 0.685 ($\pm 1.949$) & \textbf{0.321} ($\pm 0.308$)\\ 
\bottomrule
\end{tabular}
%\end{sidewaystable}
\end{table}

The CapnoBase consists of relatively clean signals collected in controlled settings, whereas the BIDMC comprises signals from critically ill patients, offering a more realistic representation of clinical conditions. Compared to the 1D convolution-based CLEP-GAN, our method demonstrates superior stability and effectiveness in handling real-world noisy conditions.

\begin{figure}[htbp!] 
\centering
\begin{subfigure}{0.45\textwidth}
  \centering
    \includegraphics[width=0.8\linewidth]{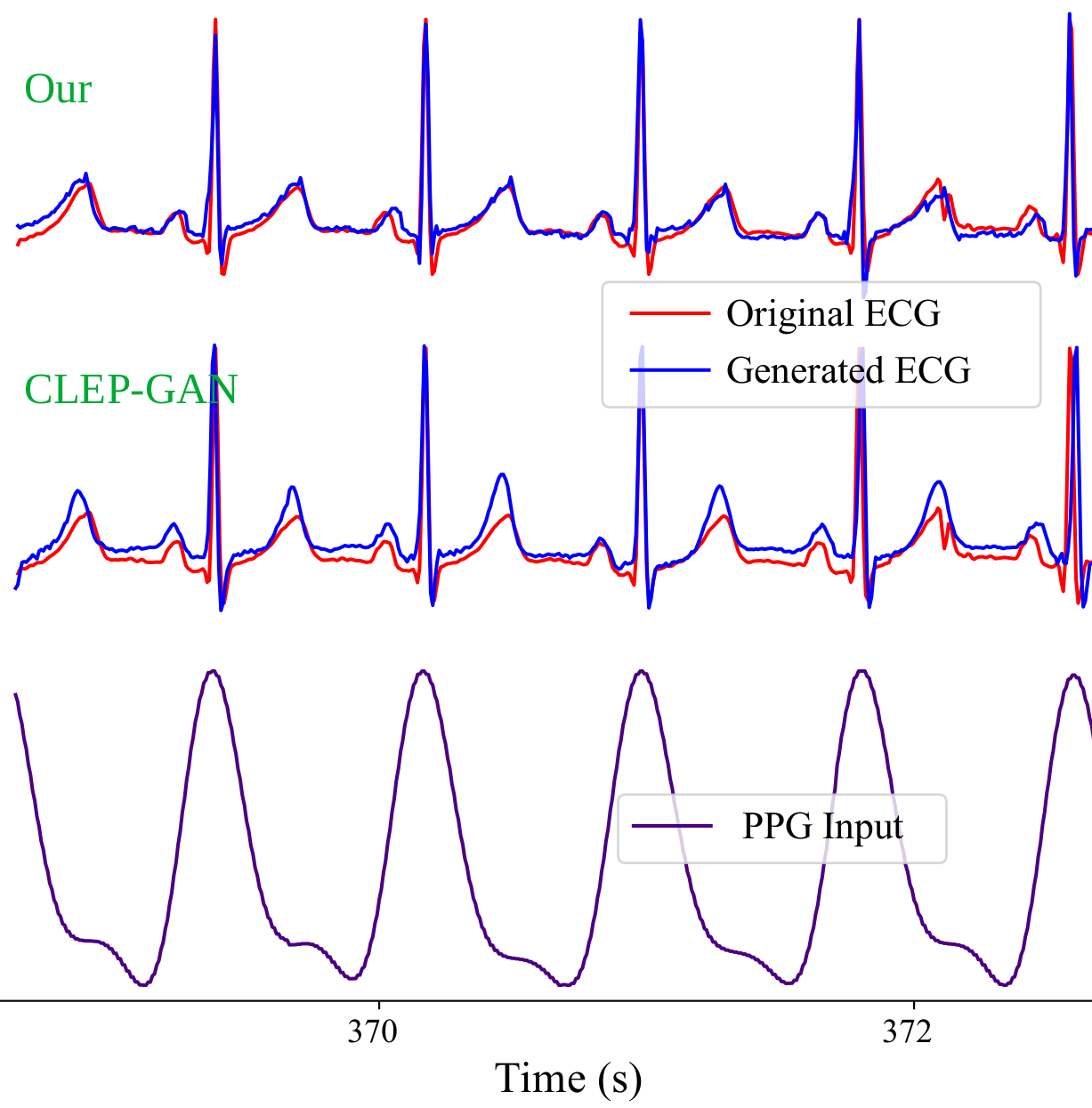}
  \caption{Generated ECG samples from record 0322 in CapnoBase Dataset.}
  \label{fig:sample_0322} 
\end{subfigure}
\hspace{0.05\textwidth} 
\begin{subfigure}{0.45\textwidth}
  \centering
    \includegraphics[width=0.8\linewidth]{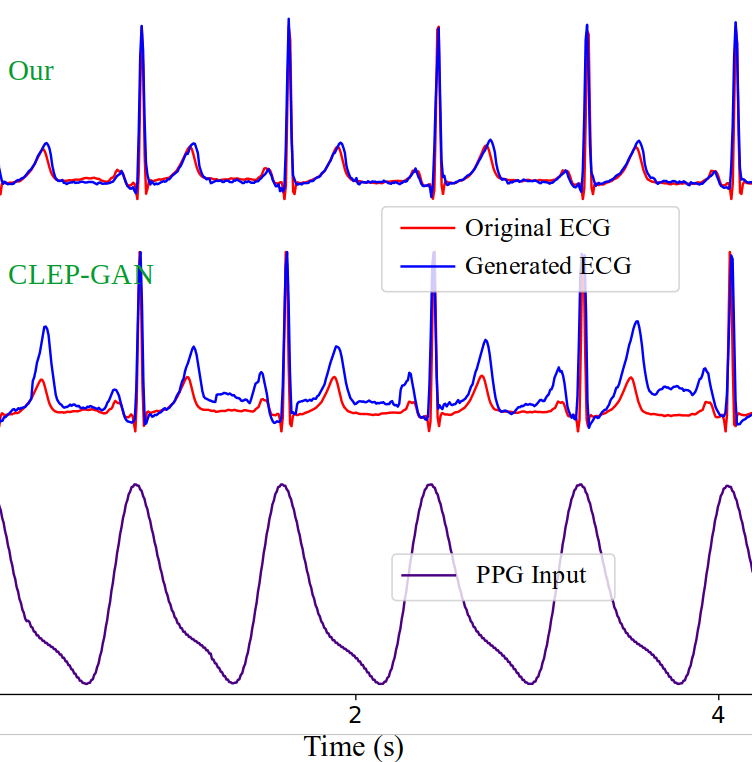}
  \caption{Generated ECG samples from record 0325 in CapnoBase Dataset.}
  \label{fig:sample_0325} 
\end{subfigure}
\caption{Comparison of ECG samples generated by our proposed method and the CLEP-GAN method.}
\label{fig:comparison_0322_0325}.
\vspace{-0.3in}
\end{figure}

Figure \ref{fig:comparison_0322_0325} provides a visual comparison of the reconstructed ECG samples generated by our method and CLEP-GAN. In both examples, our method achieves markedly improved reconstruction, yielding signals that more closely resemble the original compared to those produced by CLEP-GAN. Additional reconstruction examples are provided in \ref{appendix:extra_results}.

\section{Discussion}
\label{sec:discussion}
To evaluate the effectiveness of our four-channel image representation-based approach, we conducted comprehensive experiments addressing several key questions: (1) why ViT are more suitable than traditional Transformers for PPG-to-ECG reconstruction; (2) why padding shorter beats with values from their subsequent beats is preferable to zero-padding; and (3) why four-channel representations cannot be directly applied to existing convolution-based models. We further analyze and compare three variants of our proposed method: (i) Single-Channel Output, (ii) Four-Channel Output with convolution-based channel combination, and (iii) Four-Channel Output with weighted-sum channel combination. Additionally, for the two Four-Channel Output variants, we compare two training scenarios: applying a single MSE loss to the final ECG output versus combining it with an auxiliary MSE loss on the intermediate four-channel representation. These evaluations are followed by an ablation study.

\subsubsection{Comparison of 2D Patch-Based and 1D Patch-Based Representations}
\begin{figure}[!ht]
\vspace{-0.2cm}
\centering
\begin{subfigure}{\textwidth}
  \centering
  \begin{subfigure}{0.4\textwidth}
    \centering
    \includegraphics[width=\linewidth]{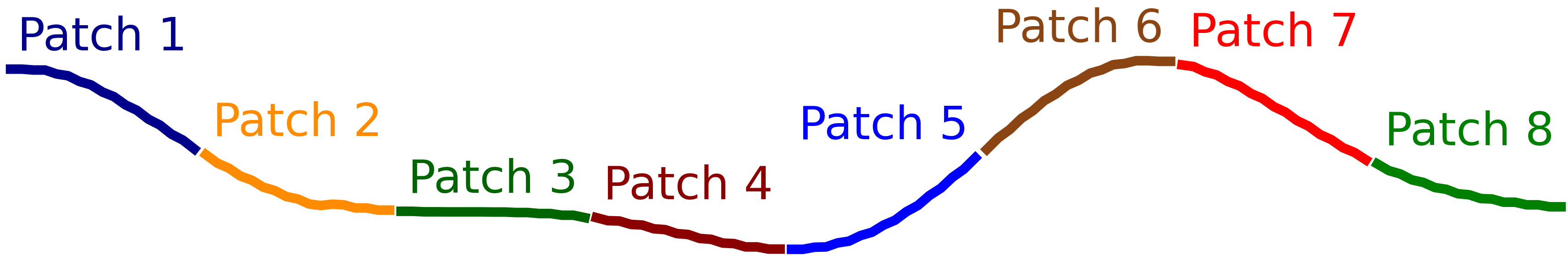}
    \caption*{(a1) The first eight patches of a 1D PPG input signal, each with a length of 16 samples.}
    \label{fig:1d_ppg_patch_signal} 
  \end{subfigure}
  \hfill
  \begin{subfigure}{0.48\textwidth}
    \centering
    \includegraphics[width=\linewidth]{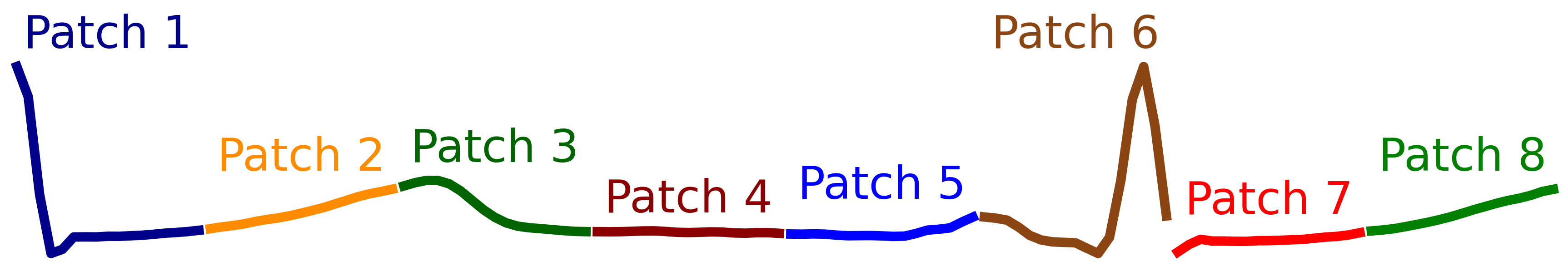}
    \caption*{(a2) The corresponding target ECG signal segments for each PPG patch.}
    \label{fig:1d_ecg_patch_signal} 
  \end{subfigure}
  \caption{1D PPG signal patches and their corresponding ECG targets.}
  \label{fig:1d_patch}
\end{subfigure}
\begin{subfigure}{\textwidth}
  \centering
  \begin{subfigure}{\textwidth}
    \centering
    \includegraphics[width=0.7\linewidth]{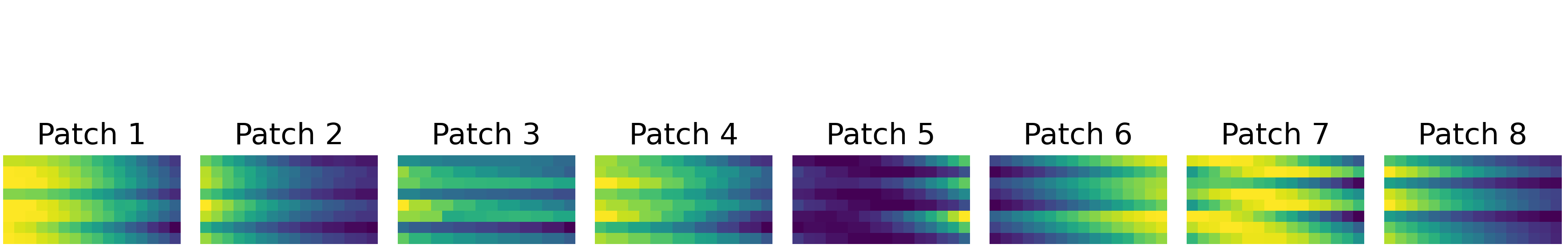}
    \caption*{(b1) The first eight patches extracted from a 2D PPG image, each with a size of \(8 \times 16\).}
    \label{fig:2d_ppg_patch_image} 
  \end{subfigure}
  \begin{subfigure}{0.48\textwidth}
    \centering
    \includegraphics[width=\linewidth]{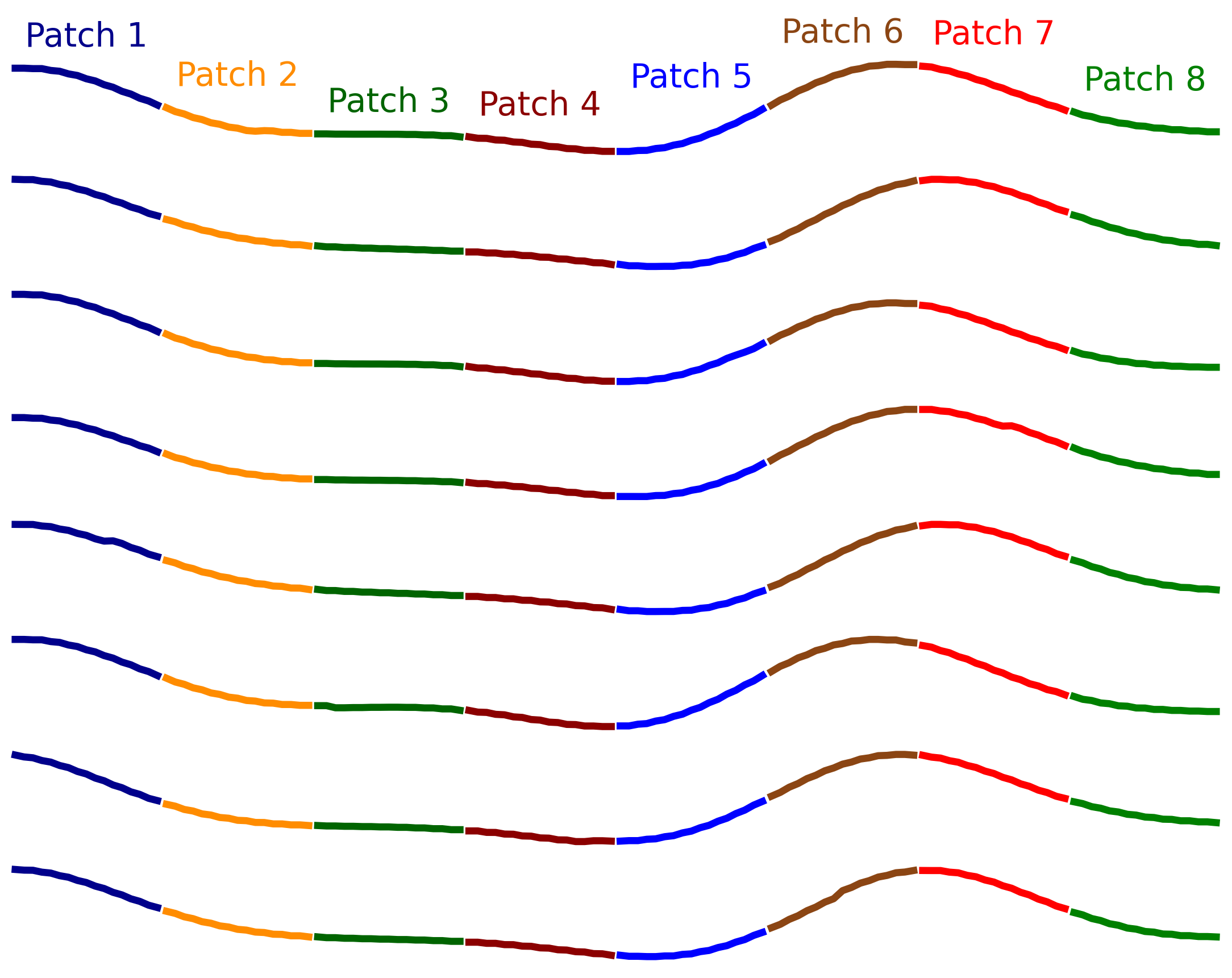}
    \caption*{(b2) Waveform representation of each 2D PPG image patch.}
    \label{fig:2d_ppg_patch_signal} 
  \end{subfigure}
  \hfill
  \begin{subfigure}{0.48\textwidth}
    \centering
    \includegraphics[width=\linewidth]{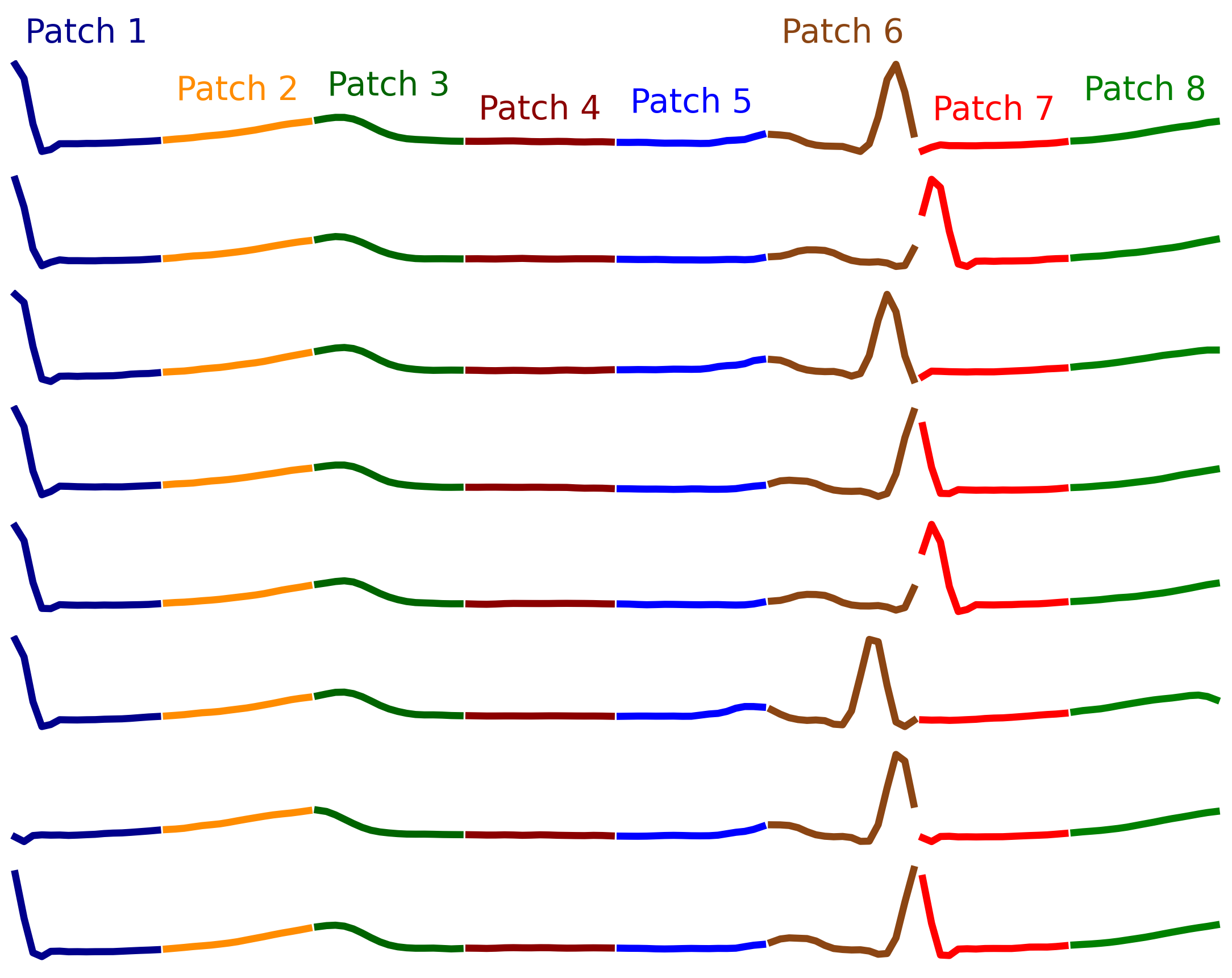}
    \caption*{(b3) The corresponding target ECG waveforms for each 2D PPG image patch.}
    \label{fig:2d_ecg_patch_signal} 
  \end{subfigure}
  \caption{2D PPG image patches, their waveforms, and corresponding ECG targets.}
  \label{fig:2d_patch}
\end{subfigure}
\caption{Visualization of the input representations used by each model: (a) 1D signal patches used in the Transformer model, and (b) 2D image patches used in the ViT.}
\label{fig:ppg_patch_ba}
% \vspace{-0.2cm}
\end{figure}

We conducted a comparative analysis to assess the conditions under which the ViT with 2D patching outperforms the 1D patch-based Transformer in ECG reconstruction. Although both models share the same encoder-decoder architecture, they differ in their input representations. The ViT processes four-channel 2D patches, whereas the Transformer operates on four-channel 1D patches. In this configuration, the 1D sequences are segmented into sequential patches along the time axis. Figure \ref{fig:ppg_patch_ba} illustrates the first eight patches extracted from a 2D PPG image and a 1D PPG signal, as used by the ViT and the Transformer networks, respectively. For better visualization, only the first channel of the image or signal is shown. In a 2D PPG image, each row represents a padded beat. When the patch size is set to \(8 \times 16 \), each patch encompasses 16 consecutive points across 8 beats. Having multiple beats within each patch allows the ViT to effectively capture variations within similar segments in different cardiac cycles (as shown in Figure \ref{fig:2d_patch}). In contrast, with 1D patching (as shown in Figure \ref{fig:1d_patch}), each patch corresponds to a single segment from one beat, which limits the model's capacity to capture inter-beat variations.

\begin{figure}[htbp!]
\centering
\begin{subfigure}{0.45\textwidth}
  \centering
  \begin{subfigure}{\textwidth}
    \centering
    \includegraphics[width=\linewidth]{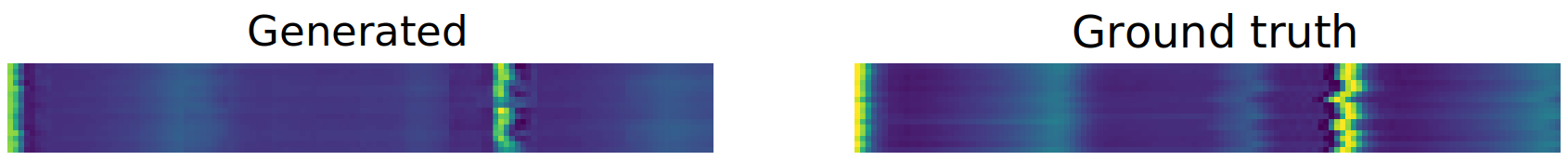}
    \caption*{(a1) Predicted ECG image using the proposed method.}
    \label{fig:vit_image42_window6} 
  \end{subfigure}
  \begin{subfigure}{\textwidth}
    \centering
    \includegraphics[width=\linewidth]{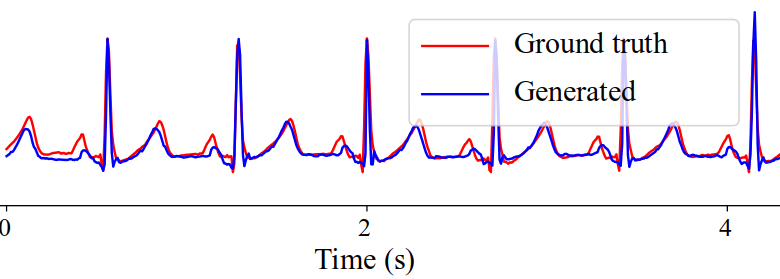}
    \caption*{(a2) Predicted ECG signal corresponding to the image in (a1).}
    \label{fig:vit_signal42_window6} 
  \end{subfigure}
  \caption{Predictions generated using the proposed method.}
  \label{fig:vit_42_window6}
\end{subfigure}
  \hfill
\begin{subfigure}{0.45\textwidth}
    \centering
    \includegraphics[width=\linewidth]{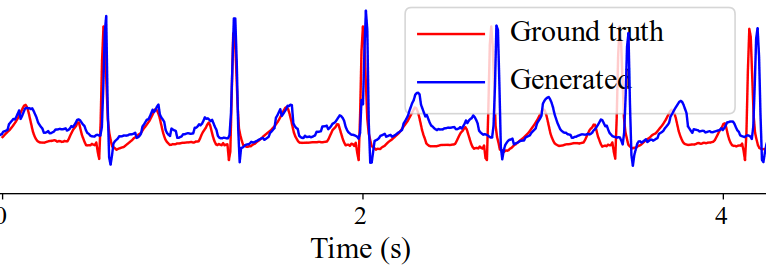}
    \vspace{1em} 
  \caption{Predictions generated using the 1D patch-based Transformer.}
  \label{fig:1dtransformer_42_window6}
\end{subfigure}
\caption{Comparison of predictions between the proposed method and the 1D patch-based Transformer for ECG reconstruction on testing record 42. Subfigures (a1) and (a2) depict the predicted ECG image and corresponding signal using the proposed method, while (b) illustrates the predicted ECG signal using the 1D patch-based Transformer.}
\label{fig:signal42_window26}
\vspace{-0.2in}
\end{figure}

As shown in Figure \ref{fig:signal_image_2}, the patches are flattened and sequentially arranged before being projected, along with position embeddings, into the transformer encoder. In our approach, the patch size is set to half the image height (i.e., the number of rows). Theoretically, larger patches contain richer information as they span multiple beats. However, using a patch size smaller than the total number of beats in an image allows the model to capture not only intra-beat but inter-beat relationships. Additionally, we observed that larger patch sizes reduce the total number of patches, leading to less stable outcomes. Figure \ref{fig:signal42_window26} compares ECG reconstructions generated by our proposed ViT-based method and the 1D patch-based Transformer. The results demonstrate that our method produces ECG signals significantly closer to the ground truth. This improvement is attributed to the ability of 2D patches to capture variations in ECG cycles, an essential capability that the 1D sequence-based Transformer lacks.

\subsubsection{Comparison of Signal Image Representations}
To generate signal images for the ViT model, we propose a beat-aligned padding method, where shorter beats are padded using values from subsequent beats. Specifically, we first determine the longest beat across two datasets and use its length as the reference. We then segment each signal into individual beats, padding shorter beats with values from their subsequent beats. This segmentation and padding process is applied consistently across all four channels, resulting in a four-channel image. 
% In our experiments, each image contains 16 beats, with a patch size of $8 \times 8$ for ViT. Increasing the number of beats per signal image while using a larger patch size could enable ViT to capture richer information within each patch. However, this approach reduces the amount of training data, which may lead to unstable performance.

\begin{figure}[htbp!] 
\centering
    \includegraphics[width=\linewidth]{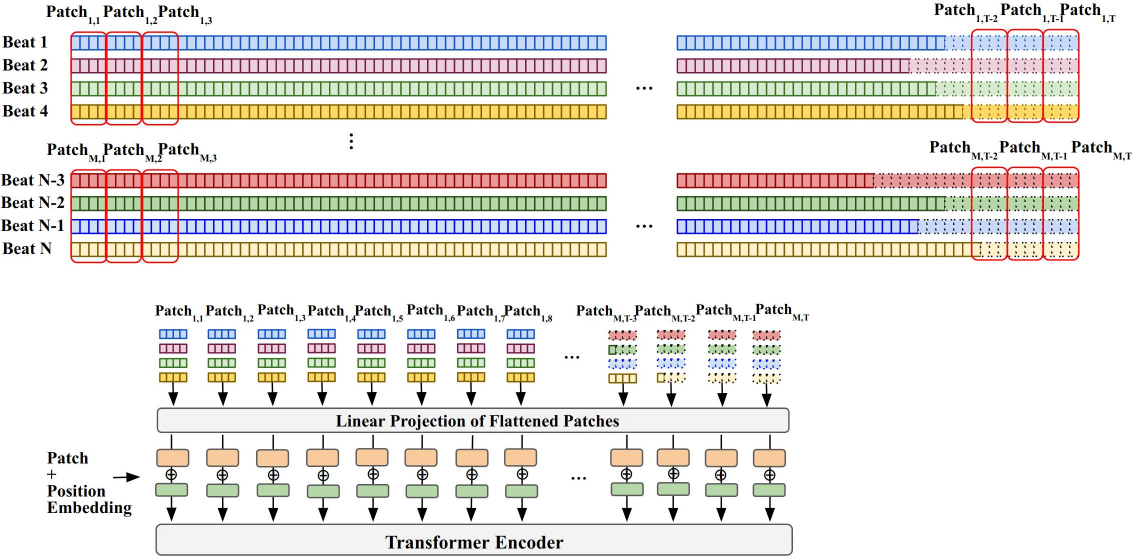}
 \caption{Patchification process for beat-aligned 2D signal images. Each row represents a single beat with four channels. Shorter beats are padded with values from subsequent beats to maintain alignment. Here, $N$ denotes the total number of beats, $M = N / 4 $ is the number of beat groups (along the vertical axis), and $T = \text{length} / 4 $ is the number of patches along the temporal axis, based on a patch size of $(4, 4)$.}
  \label{fig:signal_image_2} 
\end{figure}

This method ensures that during patchification, each patch contains a structured and representative portion of the cardiac cycle while preserving beat-to-beat variations. As a result, the model is better equipped to extract informative waveform representations and distinguish subtle differences in specific ECG components. Figure \ref{fig:signal_image_2} illustrates the patchification process of the beat-aligned method for the ViT encoder. After transforming the images into patches of size (e.g., \(4 \times 4\)), the patches are arranged sequentially, enabling the self-attention mechanism to model both intra-beat relationships and inter-beat dependencies effectively.  
We explored two alternative methods for constructing signal images. The first method, direct reshaping, transforms the 1D sequence into a 2D matrix, which serves as the input image for the ViT model. In this approach, each channel is segmented into contiguous, non-overlapping segments of equal length, which are then arranged as rows within the matrix. 
% Figure \ref{fig:signal_image_1} in Appendix \ref{sec:DR_image} illustrates this patchification process. 
The second method involves aligning each beat and padding shorter beats with zeros to standardize the length of each segment.

Considering that a typical human heart rate rarely exceeds 220 BPM for an extended period and given the signal's sample rate of 125 Hz, each cardiac cycle generally consists of more than 34 sample points. For the direct reshaping method, we set the segment length to 32 sample points, ensuring that each row of the resulting image typically contains at most one cardiac cycle. For the beat-aligned zero-padding method, the process remains similar to our proposed approach, except that shorter beats are padded with zeros instead of values from adjacent beats.

Table \ref{tab:signal_images} presents a performance comparison among the three signal image representations. In this experiment, the proposed beat-aligned padding with subsequent beat values (BA-SB) outperforms the beat-aligned padding with zeros (BA-Z). One possible explanation for the inferior performance of BA-Z is the introduction of zero-padding, which may disrupt the self-attention mechanism. During patchification, some patches contain a mix of valid signal values and padded zeros, leading to inconsistencies in feature extraction. These inconsistencies can hinder the model's ability to learn meaningful dependencies, ultimately affecting performance.

Conversely, the direct reshaping (DR) method performs significantly worse in terms of PRD, RMSE, and RT amplitude difference, exhibiting particularly high RT amplitude difference error and more than double the PRD compared to BA-SB. Despite its poor performance in these metrics, DR achieves relatively strong results for PR interval error and RT interval error. These findings suggest that the beat-aligned approach provides more stable and reliable performance, with the BA-SB method further enhancing reconstruction accuracy.

\begin{table}[htbp!]
\centering
\footnotesize
\caption{Comparison of signal image representations: Direct Reshaping (DR), Beat-Aligned Padding with Zeros (BA-Z), and the proposed Beat-Aligned Padding with the Subsequent Beat (BA-SB). Lower values across all metrics indicate better performance. Evaluation on record 42 in the BIDMC dataset.}
\label{tab:signal_images}
\begin{tabular}{l|c|c|c|c|c|c|c}
\hline
 \textbf{Method}  & \multicolumn{7}{c}{\textbf{Metrics}} \\ 
 & PRD $\downarrow$ & RMSE $\downarrow$ & R-HRV $\downarrow$ & \(\text{RE}_{\text{QRS}}\) $\downarrow$ & \(\text{RE}_{\text{PR}}\) $\downarrow$ & \(\text{RE}_{\text{RT}}\) $\downarrow$ & \(\text{RE}_{\text{AD}}\) $\downarrow$\\ 
\hline
DR    & 53.18 & 0.30 & 0.005 & 0.192 & \textbf{0.015} & \textbf{0.074} & 0.365 \\   
BA-Z  & 31.34 & 0.21 & 0.004 & 0.197 & 0.172 & 0.121 & 0.034 \\ 
BA-SB & \textbf{25.38} & \textbf{0.18} & \textbf{0.002} & \textbf{0.101} & 0.102 & 0.100 & \textbf{0.011} \\ 
\hline
\end{tabular}
\vspace{-0.1in}
\end{table}

\subsubsection{Four-Channel Approach for 1D Convolution-Based Methods} \label{sec:four_channel_1D}
To assess whether incorporating additional time sequences: PPG's first-order difference, second-order difference, and AUC, can enhance the performance of 1D convolution-based methods such as CLEP-GAN, we conducted experiments in which CLEP-GAN was trained using four-channel input sequences. In this setting, each of the four channels is represented as a 1D sequence rather than a 2D image, as used in ViT. Table \ref{tab:clep-gan_multiple_channel} presents the evaluation results when CLEP-GAN is trained on four-channel signal inputs. The findings indicate that utilizing four-channel data leads to a decline in performance compared to using only the original single-channel PPG signal across most evaluation metrics.

One possible explanation for this outcome is that Attention U-Net, the backbone of CLEP-GAN, processes signals sequentially. Although 1D convolutional models can aggregate multi-channel information using shared filters, they may struggle to effectively capture relationships between the original PPG signal and its derivative-based transformations. If the model architecture or training process does not adequately represent these dependencies, additional channels, such as the first-order difference, second-order difference, and AUC, may introduce inconsistencies rather than meaningful complementary information, ultimately limiting their contribution to improved performance. Furthermore, the limited receptive field in 1D convolution restricts the model's ability to capture long-range dependencies across channels, further reducing its capacity to leverage multi-channel data effectively.

In contrast, the ViT processes the input as a holistic representation using 2D images, enabling it to capture both temporal dependencies and structured intra-beat variations. Its self-attention mechanism allows the model to attend to all parts of the input simultaneously, effectively learning both inter-beat and intra-beat relationships for improved ECG reconstruction.

\begin{table}[htbp!]
\centering
\footnotesize
\caption{Performance of CLEP-GAN on multi-channel inputs. Two records were tested from each dataset: Record 08 from the BIDMC dataset and Record 0332 from the CapnoBase dataset. For single-channel inputs, CLEP-GAN uses only the PPG signal. For four-channel inputs, CLEP-GAN processes four time sequences: the PPG signal, its first-order difference, its second-order difference, and its AUC.}
\label{tab:clep-gan_multiple_channel}
\begin{tabular}{l|c|c|c|c|c}
\hline
    &    & \multicolumn{4}{c}{\textbf{Metrics}} \\ 
\textbf{Record} & \textbf{Input} & PRD $\downarrow$ & RMSE $\downarrow$ & R-HRV$\downarrow$ & \(\text{RE}_{\text{QRS}}\) $\downarrow$ \\ 
\hline
 & PPG & \textbf{51.75} &  \textbf{0.34} &  \textbf{0.044} &  \textbf{0.015}\\  
08 & Four-Channel  & 67.68 & 0.45 & 0.090 & 0.018 \\       
\hline
 & PPG   & \textbf{46.72} & 0.31 & \textbf{0.000} & \textbf{0.055}\\
0332 & Four-Channel & 58.72 & \textbf{0.30} & 0.004 & 0.287\\     
\bottomrule
\end{tabular}
\vspace{-0.1in}
\end{table}

\subsubsection{Analysis of Output Image Combination Methods}
In our proposed method, we introduce two approaches for predicting ECG images. The first approach specifies the ViT decoder to output a single-channel ECG image. The second approach sets the number of output channels to four, corresponding to the input channels. These four channels represent the 2D ECG image, its first-order difference image, second-order difference image, and its AUC image (as illustrated in Figure \ref{fig:four_channel_signal_0325} in the Appendix). In this multi-channel approach, the final ECG image is reconstructed by combining these four channels. 

To combine the four channels, we propose two methods. The convolution-based combination method utilizes a learnable 2D convolution layer to merge the four channels. The weighted sum combination method applies a weighted summation of the channels, where the weights are learned during training.  

Table \ref{tab:output_combination} summarizes the performance of these approaches, highlighting their respective strengths based on different evaluation metrics. The single-channel output method achieves the lowest RMSE and \(\text{RE}_{\text{QRS}}\). The convolution-based combination method demonstrates superior performance in \(\text{RE}_{\text{PR}}\), \(\text{RE}_{\text{RT}}\), and \(\text{RE}_{\text{AD}}\). Although the weighted sum combination method generally underperforms compared to the other two methods, it achieves the lowest HRV error.

\begin{table}[htbp!]
\centering
\footnotesize
\caption{Comparison of output combination approaches, convolution-based combination (Conv) and weighted sum combination (WS), against the single-channel output method. The performance is evaluated on four records: 0322 and 0325 from the CapnoBase dataset, and 22 and 42 from the BIDMC dataset. The table reports the average results across all records, with lower values indicating better performance across all metrics.}
\label{tab:output_combination}
\begin{tabular}{l|c|c|c|c|c|c|c}
\hline
 & \multicolumn{4}{c}{\textbf{Metrics}} \\ 
\textbf{Output} & PRD $\downarrow$ & RMSE $\downarrow$ & R-HRV $\downarrow$ & \(\text{RE}_{\text{QRS}}\) $\downarrow$ & \(\text{RE}_{\text{PR}}\) $\downarrow$ & \(\text{RE}_{\text{RT}}\) $\downarrow$ & \(\text{RE}_{\text{AD}}\) $\downarrow$ \\ 
\hline
 Single & 26.56 & \textbf{0.18} & 0.001 &  \textbf{0.113} & 0.092 & 0.175 & 0.095\\ 
 Conv   & \textbf{26.39} &  \textbf{0.18} & 0.004 & 0.157 &  \textbf{0.089} &  \textbf{0.115} &  \textbf{0.086}\\  
 WS     & 32.03 & 0.21 &  \textbf{0.000} & 0.253 & 0.106 & 0.186 & 0.124\\
\bottomrule
\end{tabular}
%\end{sidewaystable}
\vspace{-0.2in}
\end{table}

\subsubsection{Experiments on Loss Functions}  
% \paragraph{Impact of QRS Complex-Enhanced Loss on ECG Image Generation}
% In our proposed methods, we use two loss functions: reconstruction loss (\(L_{R_{1C}}\)) and QRS complex-enhanced loss (\(L_{QRS}\)), as described in Section \ref{sec:objective}. In this section, we first evaluate the impact of the QRS complex-enhanced loss on the generation results.  

% Table \ref{tab:mse_qrs_losses} presents a comparison between using \(L_{QRS}\) alone and the combined loss of \(L_{R_{1C}}\) and \(L_{QRS}\). The results show that incorporating QRS complex-enhanced loss leads to lower HRV errors, which can be attributed to the improved reconstruction of R peaks. Additionally, in this example, incorporating \(L_{QRS}\) also helps reduce notable PRD and RMSE.

% \begin{table}[h!]
% \centering
% \caption{The proposed single-channel output method evaluated on record 0332 from the CapnoBase dataset.}
% \label{tab:mse_qrs_losses}
% \renewcommand{\arraystretch}{1.2}  
% \setlength{\tabcolsep}{1.5pt}  
% \begin{tabular}{l|c|c|c|c|c|c|c}
% \hline
%  & \multicolumn{7}{c}{\textbf{Metrics}} \\ 
% \textbf{Loss} & PRD & RMSE & R-HRV & QRS Area & PR Inte. & RT Inte. & RT Ampl. \\ 
% \hline
%  $L_{R_{1C}}$              & 21.57 & 0.15 & 0.002 & 0.097 & \textbf{0.124} & \textbf{0.072} & \textbf{0.061} \\
% $L_{R_{1C}} + L_{QRS}$     & \textbf{18.14} & \textbf{0.13} & \textbf{0.000} & \textbf{0.096} & 0.136 & 0.074 & 0.085 \\ 
% \bottomrule
% \end{tabular}
% \end{table}

\paragraph{Impact of QRS Complex-Enhanced Loss}  
Our proposed method incorporates two loss functions: the reconstruction loss (\(L_{R_{1C}}\)) and the QRS complex-enhanced loss (\(L_{QRS}\)), as described in Section \ref{sec:objective}. This section evaluates the effect of the QRS complex-enhanced loss on ECG generation.

Table \ref{tab:mse_qrs_losses} presents a performance comparison between using \(L_{QRS}\) alone and the combined loss \(L_{R_{1C}} + L_{\text{QRS}}\). The results demonstrate that incorporating \(L_{\text{QRS}}\) effectively reduces HRV errors, which is attributed to improved reconstruction of R peaks. Furthermore, adding \(L_{\text{QRS}}\) leads to a significant reduction in PRD and RMSE.

\begin{table}[htbp!]
\centering
\footnotesize
\caption{Evaluation of QRS Complex-Enhanced Loss. The experiment was conducted on record 0332 from the CapnoBase dataset.}
\label{tab:mse_qrs_losses}
\begin{tabular}{l|c|c|c|c|c|c|c}
\hline
 & \multicolumn{7}{c}{\textbf{Metrics}} \\ 
\textbf{Loss} & PRD$\downarrow$ & RMSE$\downarrow$ & R-HRV$\downarrow$ & \(\text{RE}_{\text{QRS}}\)$\downarrow$ & \(\text{RE}_{\text{PR}}\)$\downarrow$ & \(\text{RE}_{\text{RT}}\)$\downarrow$ & \(\text{RE}_{\text{AD}}\)$\downarrow$\\ 
\hline
 $L_{R_{1C}}$              & 21.57 & 0.15 & 0.002 & 0.097 & \textbf{0.124} & \textbf{0.072} & \textbf{0.061} \\
$L_{R_{1C}} + L_{QRS}$     & \textbf{18.14} & \textbf{0.13} & \textbf{0.000} & \textbf{0.096} & 0.136 & 0.074 & 0.085 \\ 
\bottomrule
\end{tabular}
\vspace{-0.2cm}
\end{table}

\paragraph{Impact of Full-Channel Reconstruction Loss}
For the four-channel output approach, we evaluate whether incorporating the reconstruction loss of the full-channel output enhances performance. We compare two loss configurations: (1) using only the reconstruction loss for the final ECG image, denoted as \(L_{R_{1C}}\), and (2) incorporating both the reconstruction loss of the final single-channel ECG image and the reconstruction loss of the intermediate four-channel output (i.e., the four-channel ECG representation before being combined into the final image), resulting in the total loss \(L_{R_{1C}} + L_{R_{4C}}\). The reconstruction loss for the four-channel output is computed using the MSE, defined as $ L_{R_{4C}} = \frac{1}{P_4} \sum_{i=1}^{P_4} (\hat{Y}_I[i] - Y_I[i] )^2$. \(P_4\) represents the total number of pixels across all four channels. Table \ref{tab:two_lossses} presents the performance comparison between \(L_{R_{1C}}\) and \(L_{R_{1C}} + L_{R_{4C}}\). Figure \ref{fig:combined_output} in the Appendix provides a visualization of the generated ECG images using this combined loss.  

\begin{table}[htbp!]
\centering
\footnotesize
\caption{Comparison of results using two different loss functions. The evaluation is conducted on record 0325 from the CapnoBase dataset. Lower values indicate better performance across all metrics.}
\label{tab:two_lossses}
\begin{tabular}{l|c|c|c|c|c|c}
\hline
    &    & \multicolumn{4}{c}{\textbf{Metrics}} \\ 
\textbf{Output} & \textbf{Recon. Loss} & PRD $\downarrow$ & RMSE $\downarrow$ & R-HRV $\downarrow$ & \(\text{RE}_{\text{QRS}}\) $\downarrow$ & \(\text{RE}_{\text{AD}}\) $\downarrow$ \\ 
\hline
             & $L_{R_{1C}}$         & \textbf{23.85} & \textbf{0.17} & 0.002 & \textbf{0.243} & \textbf{0.169}\\
Conv  & $L_{R_{1C}} + L_{R_{4C}}$ & 26.37 & 0.18 & \textbf{0.000} & 0.312 &  0.202 \\ 
\hline
             & $L_{R_{1C}}$         & \textbf{21.56} & \textbf{0.16} & \textbf{0.000} & 0.349 & \textbf{0.102}\\
WS   & $L_{R_{1C}}+L_{R_{4C}}$  & 25.28 & 0.18 & \textbf{0.000} & \textbf{0.334} & 0.194 \\ 
\bottomrule
\end{tabular}
\end{table}

\begin{figure}[htbp!]
\centering
  \begin{subfigure}{0.45\textwidth}
      \centering
      \includegraphics[width=\linewidth]{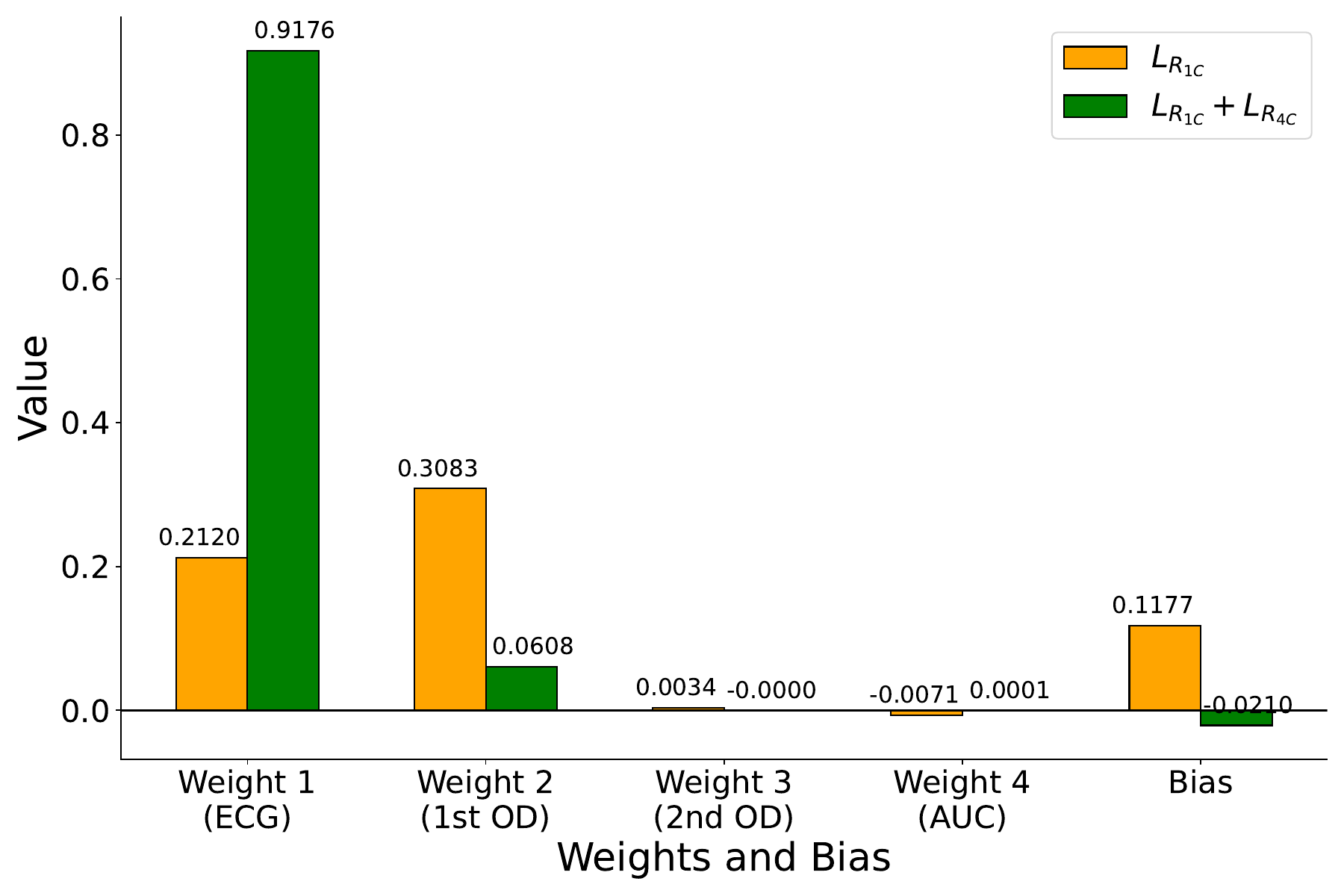}
      \caption{Visualization of weights and biases from the convolution-based combination approach.}
      \label{fig:weight_conv_0325}
  \end{subfigure}
\vspace{1em} 
\begin{subfigure}{0.45\textwidth}
  \centering
    \includegraphics[width=\linewidth]{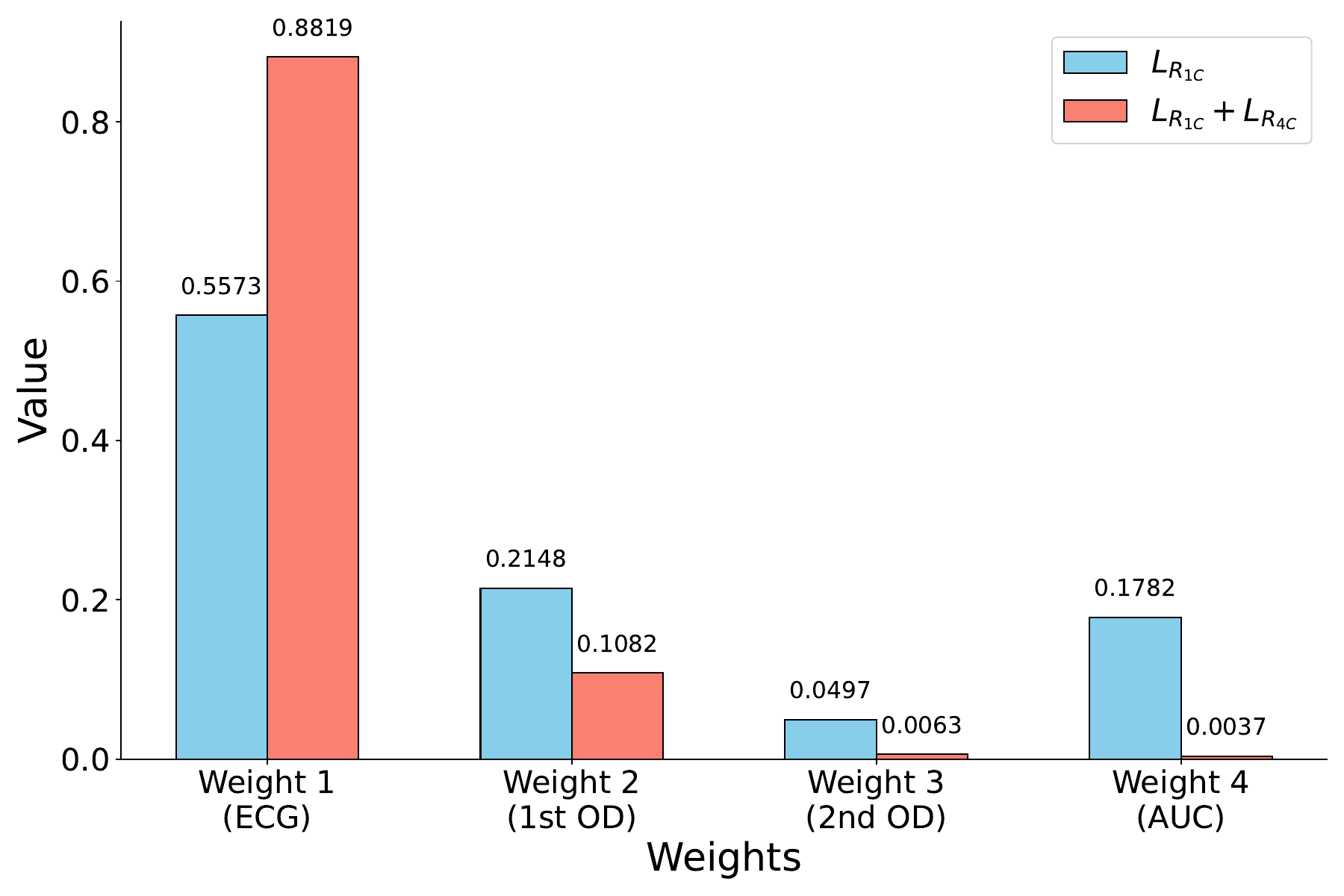}
  \caption{Visualization of weights from the weighted sum-based combination approach.}
  \label{fig:weight_ws_0325}
\end{subfigure}
\caption{Comparison of weight visualizations for two combination methods: (a) the convolution-based combination approach and (b) the weighted sum-based combination approach. Each method utilizes two types of reconstruction losses: (1) the final ECG image MSE loss (\(L_{R_{1C}}\)) and (2) the combined loss (\(L_{R_{1C}} + L_{R_{4C}}\)), where \(L_{R_{4C}}\) represents the MSE loss for the four-channel intermediate ECG image.}
\label{fig:weight_0325}
\vspace{-0.1in}
\end{figure}

From Table \ref{tab:two_lossses}, we observe that incorporating \(L_{R_{4C}}\) leads to an increase in PRD, RMSE, and \(\text{RE}_{\text{AD}}\) for both combination approaches. This suggests that adding the four-channel reconstruction loss introduces trade-offs that may affect the quality of the final ECG signal. Although incorporating \(L_{R_{4C}}\) reduces reconstruction errors in the three additional channels (first-order difference, second-order difference, and AUC), which is expected to improve the waveform representation, it also shifts the model's focus. As shown in Figure \ref{fig:weight_0325}, the inclusion of \(L_{R_{4C}}\) results in a greater weight being assigned to the ECG channel, while the contributions of the other three channels decrease. This happens because the ECG channel directly corresponds to the final ECG image, prompting the model to prioritize it more heavily when the four-channel reconstruction loss is included.

\subsubsection{Ablation Study on Input Channels for ECG Prediction}
In this ablation study, we examine the impact of three additional input channels, the first-order and second-order differences of PPG, as well as its AUC, on ECG prediction when using PPG as the primary input. The experiments were conducted using our single-channel output approach, and the results are presented in Table \ref{tab:ablation}. When the model uses PPG and its first-order difference (1st.) as input, it achieves improved results in \(\text{RE}_{\text{QRS}}\) and \(\text{RE}_{\text{RT}}\). Adding the second-order difference (2nd.) further reduces \(\text{RE}_{\text{PR}}\) and \(\text{RE}_{\text{AD}}\); however, it also increases PRD and RMSE compared to using only PPG and its first-order difference.

When using PPG and its AUC as input, we observe a notable reduction in PRD, RMSE, and \(\text{RE}_{\text{PR}}\), while \(\text{RE}_{\text{RT}}\) and \(\text{RE}_{\text{AD}}\) show a slight increase compared to using PPG alone. Combining PPG, first-order difference, and AUC as inputs yields the lowest \(\text{RE}_{\text{RT}}\) but results in higher PRD and RMSE.

Finally, when all four channels: PPG, first-order difference, second-order difference, and AUC, are included, the model effectively leverages the advantages of each channel, leading to a more balanced performance.

\begin{table}[htbp!]
\centering
\scriptsize
\caption{Ablation study on the effect of different input channel configurations on ECG prediction.}
\label{tab:ablation}
\begin{tabular}{l|c|c|c|c|c|c|c}
\hline
\textbf{Input} & \multicolumn{4}{c}{\textbf{Metrics}} \\ 
\textbf{Channel} & PRD$\downarrow$ & RMSE$\downarrow$ & R-HRV$\downarrow$ & \(\text{RE}_{\text{QRS}}\)$\downarrow$ & \(\text{RE}_{\text{PR}}\)$\downarrow$ & \(\text{RE}_{\text{RT}}\)$\downarrow$ & \(\text{RE}_{\text{AD}}\)$\downarrow$ \\
\hline
PPG               & 23.30 & 0.16 & \textbf{0.000} & 0.145 & 0.120 & 0.028 & 0.162\\
\hline
PPG, 1st.         & 23.02 & 0.16 & \textbf{0.000} & \textbf{0.066} & 0.140 & \textbf{0.023} & 0.168 \\
\hline
PPG,1st.,2nd. & 25.44 & 0.18 & \textbf{0.000} & 0.105 & 0.098 & 0.029 & 0.160\\ 
\hline
PPG, AUC            & \textbf{17.29} & \textbf{0.13} & \textbf{0.000} & 0.115 & \textbf{0.066} & 0.066 & 0.184\\ 
\hline
PPG,1st.,AUC        & 25.31 & 0.18 & \textbf{0.000} & 0.121 & 0.123 & \textbf{0.020} & 0.170 \\ 
\hline
Four-Channel        & 18.14 & \textbf{0.13} & \textbf{0.000} & 0.096 & 0.136 & 0.074 & \textbf{0.085}\\
\bottomrule
\end{tabular}
%\end{sidewaystable}
\vspace{-0.2cm}
\end{table}

\section{Conclusion}\label{sec:conclusion}

This study presents a novel approach for ECG reconstruction from PPG by leveraging a ViT and a four-channel image-based signal representation. Unlike traditional 1D CNN-based models, our method reformulates PPG sequences into structured 2D signal images comprising the raw waveform, its first- and second-order differences, and the AUC. This enriched multi-channel representation enables the ViT to model both intra-beat morphology and cross-cycle temporal dynamics through self-attention mechanisms. 

The proposed framework demonstrates strong empirical performance, achieving up to a 29\% reduction in PRD and a 15\% reduction in RMSE compared to state-of-the-art baselines. To preserve temporal continuity across cardiac cycles, we introduced a beat-padding strategy that maintains beat-to-beat continuity better than conventional zero-padding. Additionally, we employed clinically relevant metrics, such as QRS area and RT interval errors, to enhance interpretability beyond conventional point-wise evaluations.

Despite its strengths, the current implementation has limitations. Chief among them is the use of a fixed patch size in the ViT architecture, which may not generalize well to signals with varying sampling rates or beat durations. In high-frequency signals, large patches may span multiple waveform components, reducing the model's ability to resolve distinct features. Moreover, static patching may miss subtle yet clinically important morphological variations, particularly in datasets with high inter-subject variability.

Nonetheless, our proposed framework provides a solid foundation for developing practical algorithms for ECG reconstruction. Future work should explore adaptive or learnable patching strategies that dynamically align patch size with signal properties. Integrating auxiliary networks trained to detect specific waveform boundaries, such as the onset of the P wave or the offset of the T wave, may enhance boundary localization and improve clinical interpretability. Furthermore, extending this approach to multi-lead ECG reconstruction and evaluating its robustness under motion artifacts or across wearable PPG devices would enhance its real-world applicability. Beyond ECG reconstruction, the proposed ViT-based framework provides a generalizable solution for modeling cyclic physiological signals, such as respiratory or arterial waveforms, contributing broadly to biomedical signal processing and pattern recognition.

\appendix

\section{Visualization of Four-Channel Output}
\begin{figure} [!ht]
\vspace{-0.2cm}
\centering
  \begin{subfigure}{0.32\textwidth}
    \centering
    \includegraphics[width=\linewidth]{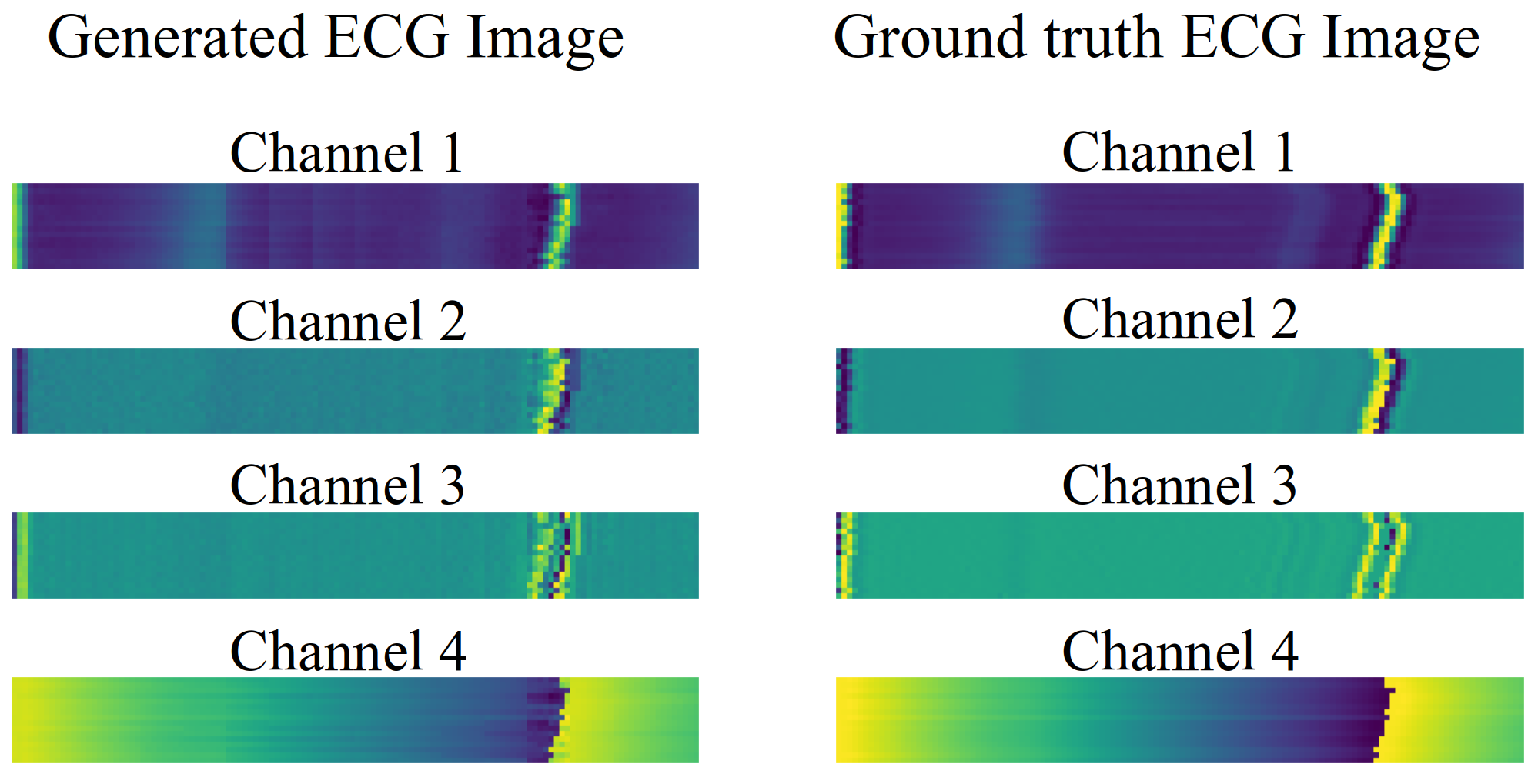}
    \caption{Generated four-channel ECG image.}
    \label{fig:four_channel_image_0325} 
  \end{subfigure}
   \begin{subfigure}{0.32\textwidth}
    \centering
    \includegraphics[width=\linewidth]{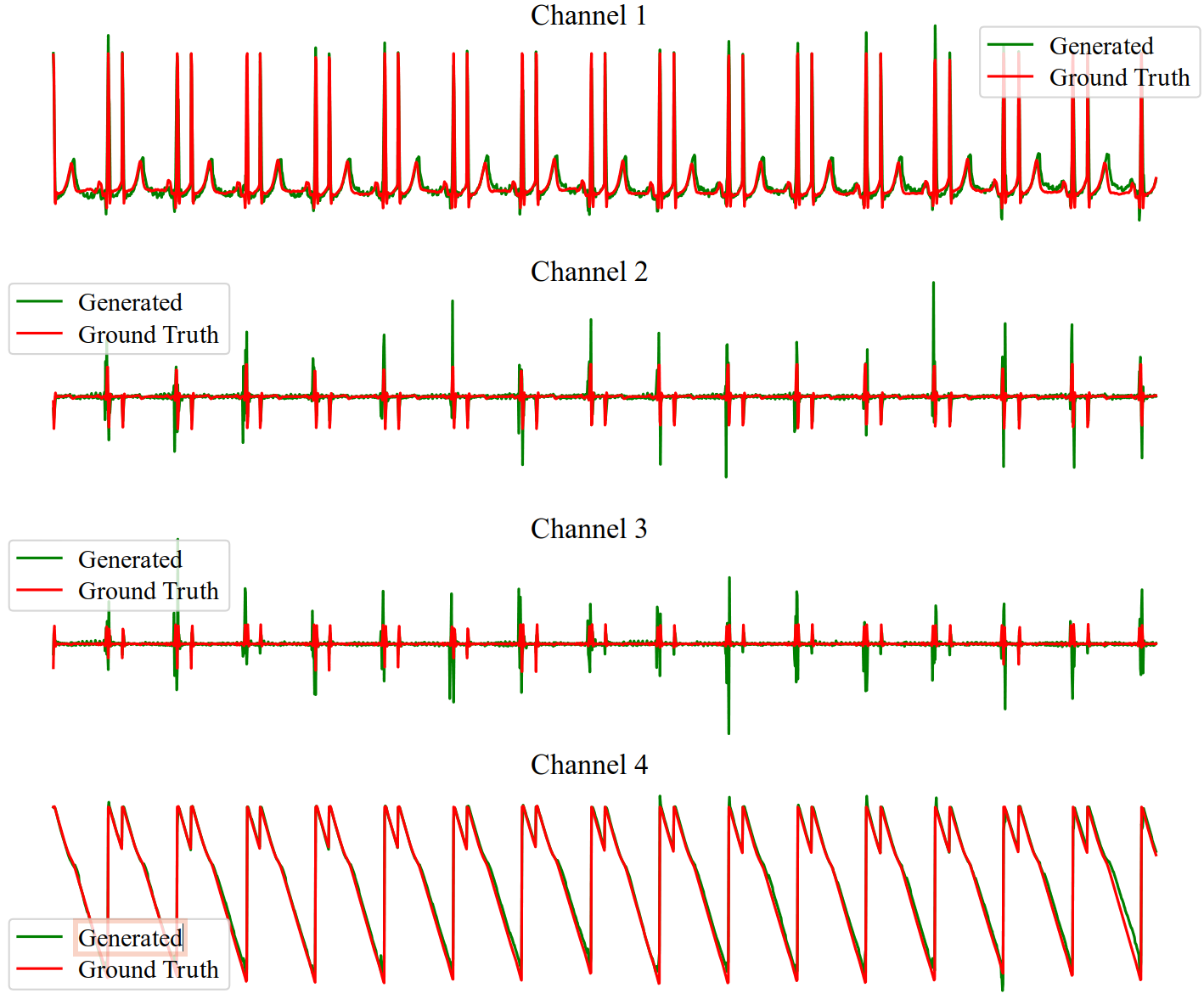}
    \caption{Corresponding four-channel signal, where each beat is padded with its subsequent beat.}
    \label{fig:four_channel_signal_0325} 
  \end{subfigure}
  \begin{subfigure}{0.32\textwidth}
    \centering
    \includegraphics[width=\linewidth]{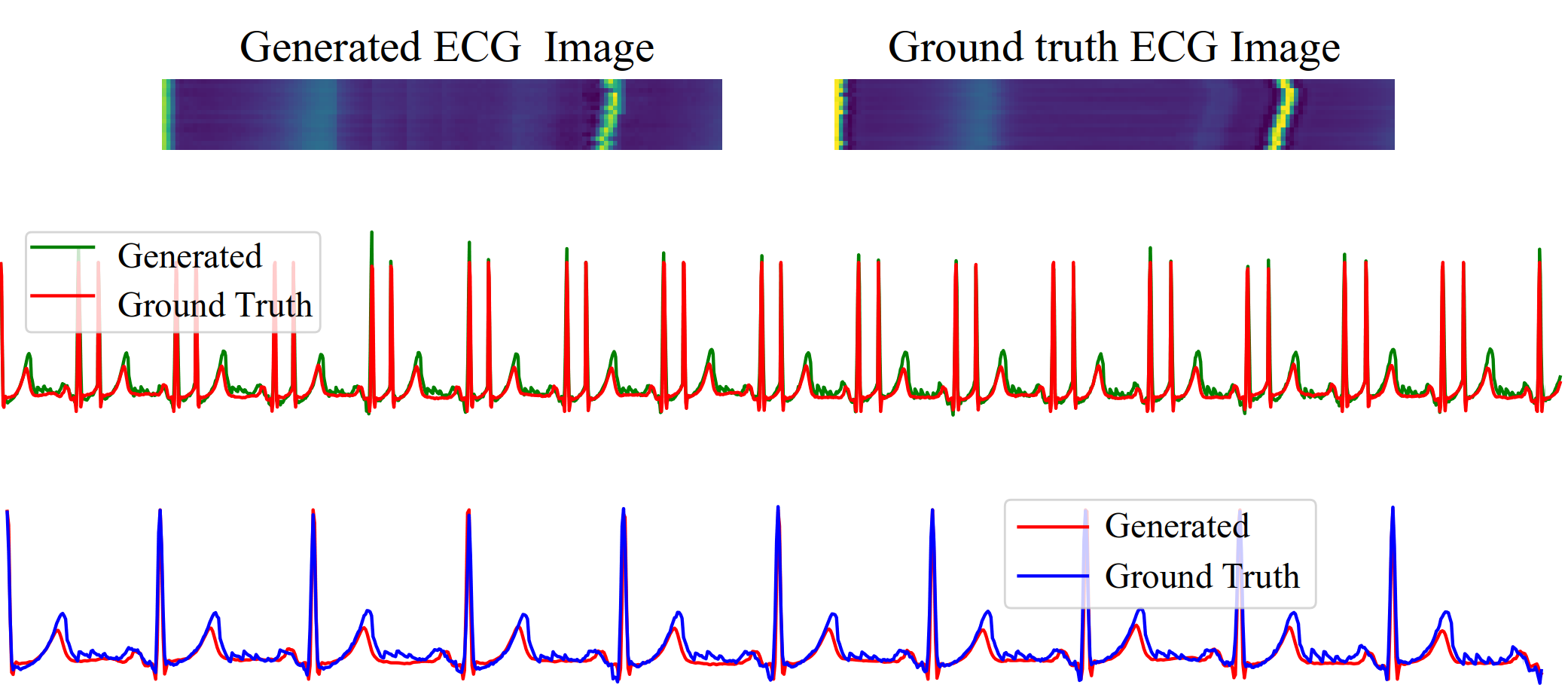}
    \caption{Final output after combination. The top-left image represents the reconstructed ECG image, the middle plot shows the corresponding ECG signal, and the bottom plot presents the final ECG signal after removing padding.}
    \label{fig:output_signal_0325} 
  \end{subfigure}
\caption{Visualization of the four-channel output and corresponding signals for record 0325 in the CapnoBase dataset. The output is generated using the weighted sum-based combination approach with two reconstruction losses ($L_{R_{1C}}$ + $L_{R_{4C}}$). Figures \ref{fig:four_channel_image_0325} and \ref{fig:four_channel_signal_0325} depict the intermediate four-channel ECG image and its corresponding signal, while Figure \ref{fig:output_signal_0325} illustrates the final reconstructed ECG signal after padding removal.}
\label{fig:combined_output}
\vspace{-0.2cm}
\end{figure}

Figure \ref{fig:combined_output} presents the four-channel output images, their corresponding signals, and the final combined output, generated using the weighted sum-based combination approach with two reconstruction losses ($L_{R_{1C}}$ + $L_{R_{4C}}$). The predicted final ECG signal consists of concatenated padded beats. To obtain the correct ECG signal, the padding must be removed, as illustrated in the bottom plot of Figure \ref{fig:output_signal_0325}.

\section{Additional ECG Signal Reconstructions}\label{appendix:extra_results}
\begin{figure}[!ht]
\centering
  \begin{subfigure}{0.38\textwidth}
    \centering
    \includegraphics[width=\linewidth]{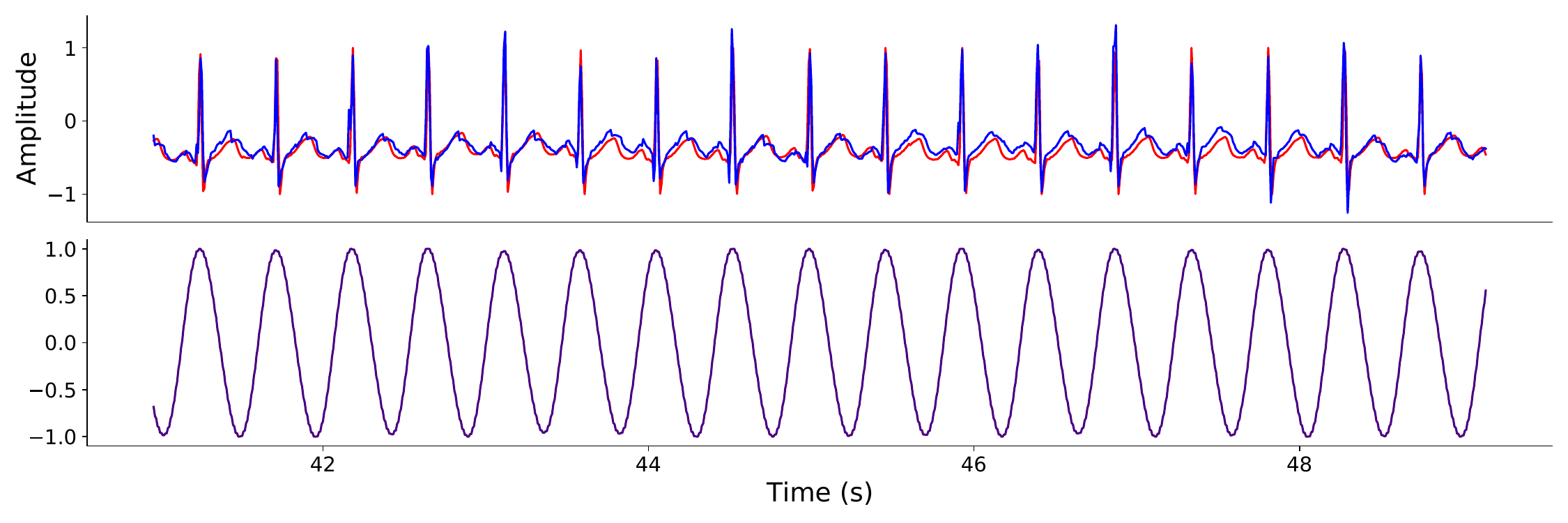}
    \caption{Record 0016.}
    \label{fig:record_0016} 
  \end{subfigure}
  \begin{subfigure}{0.38\textwidth}
    \centering
    \includegraphics[width=\linewidth]{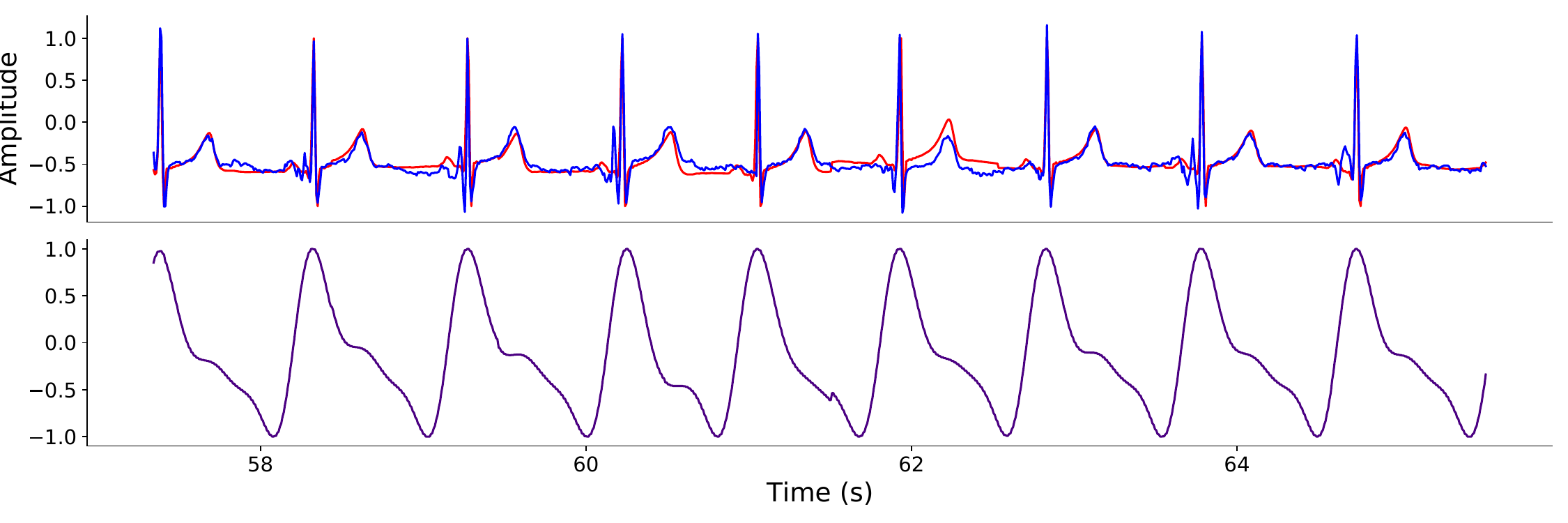}
    \caption{Record 0029.}
    \label{fig:record_0029} 
  \end{subfigure}
  \begin{subfigure}{0.38\textwidth}
    \centering
    \includegraphics[width=\linewidth]{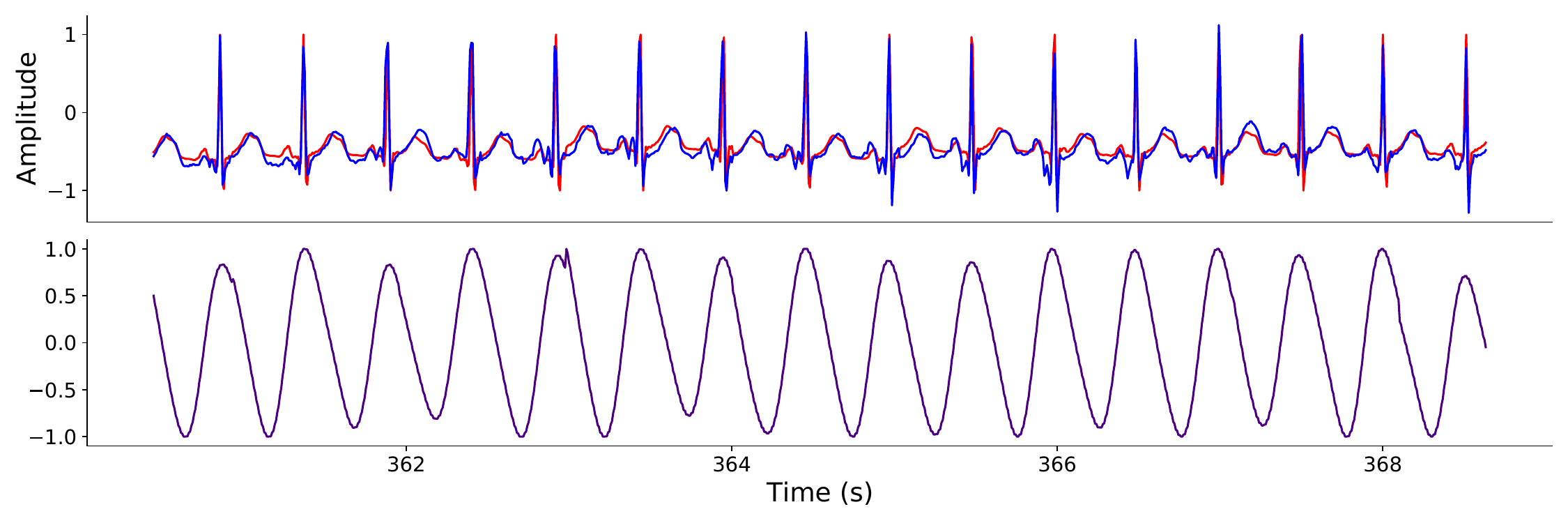}
    \caption{Record 0035.}
    \label{fig:recored_0035} 
  \end{subfigure}
  \begin{subfigure}{0.38\textwidth}
    \centering
    \includegraphics[width=\linewidth]{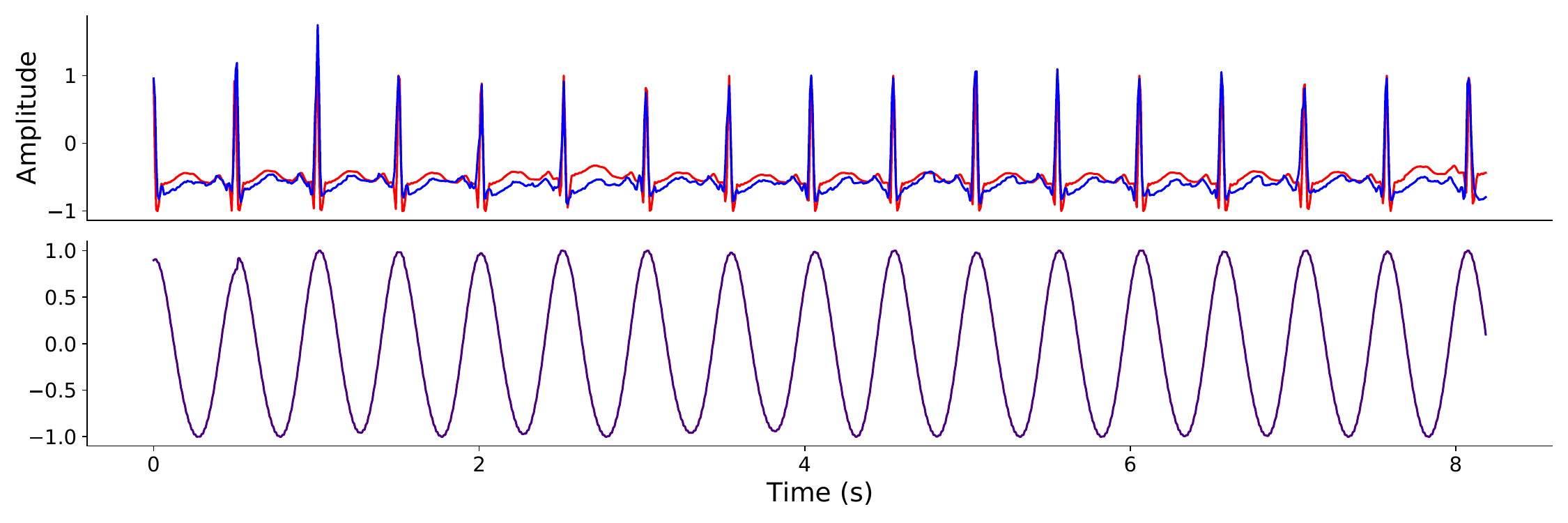}
    \caption{Record 0104.}
    \label{fig:record_0104} 
  \end{subfigure}
  \begin{subfigure}{0.38\textwidth}
    \centering
    \includegraphics[width=\linewidth]{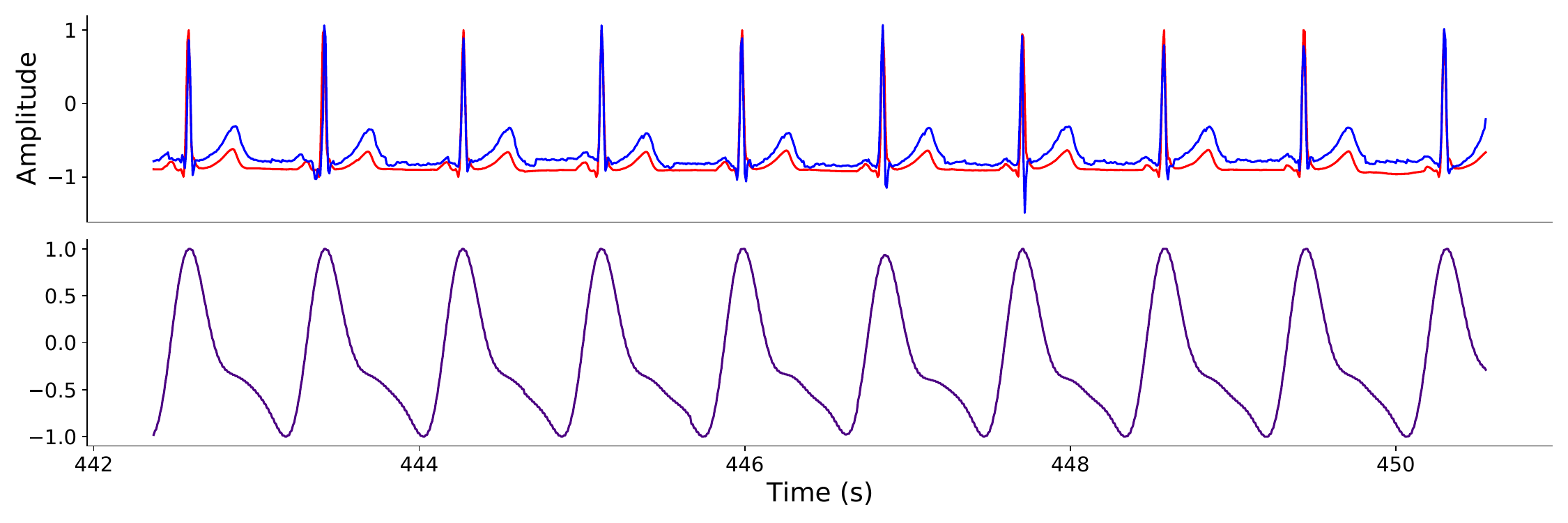}
    \caption{Record 0121.}
    \label{fig:record_0121} 
  \end{subfigure}
  % \begin{subfigure}{0.3\textwidth}
  %   \centering
  %   \includegraphics[width=\linewidth]{./figs/add_examples/epoch2000_signal0148_window36.pdf}
  %   \caption{Record 0148.}
  %   \label{fig:record_0148} 
  % \end{subfigure}
  \begin{subfigure}{0.38\textwidth}
    \centering
    \includegraphics[width=\linewidth]{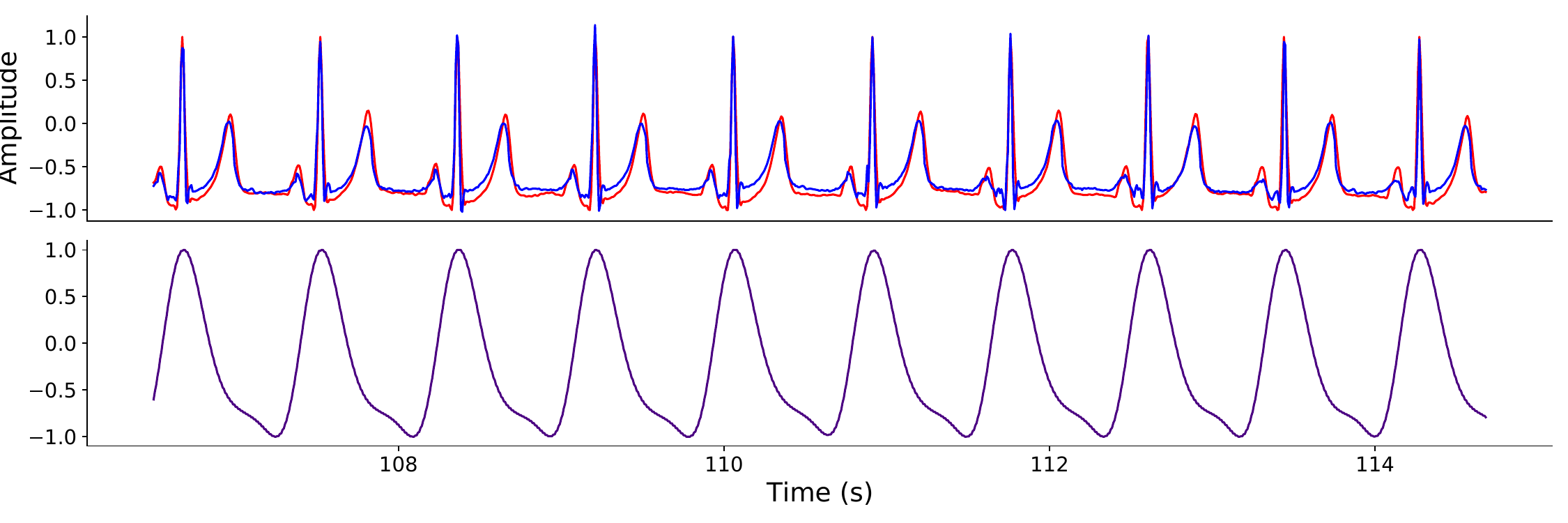}
    \caption{Record 0311.}
    \label{fig:recored_0311} 
  \end{subfigure}
\begin{subfigure}{0.38\textwidth}
    \centering
    \includegraphics[width=\linewidth]{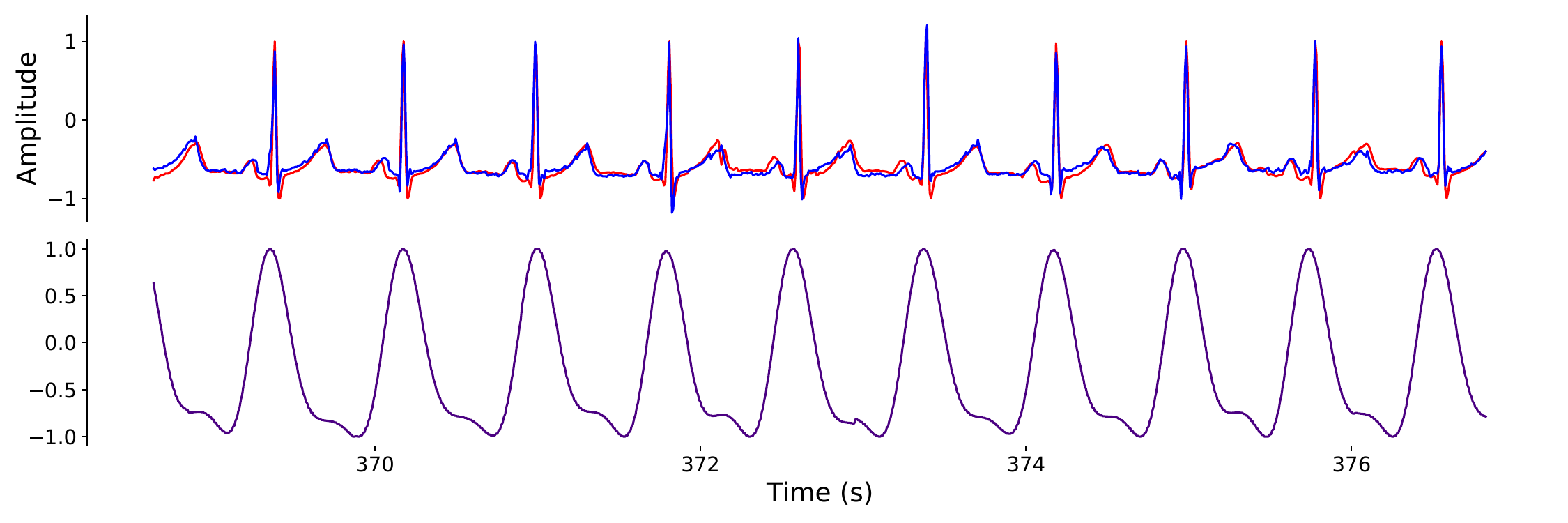}
    \caption{Record 0322.}
    \label{fig:record_0322} 
  \end{subfigure}
  \begin{subfigure}{0.38\textwidth}
    \centering
    \includegraphics[width=\linewidth]{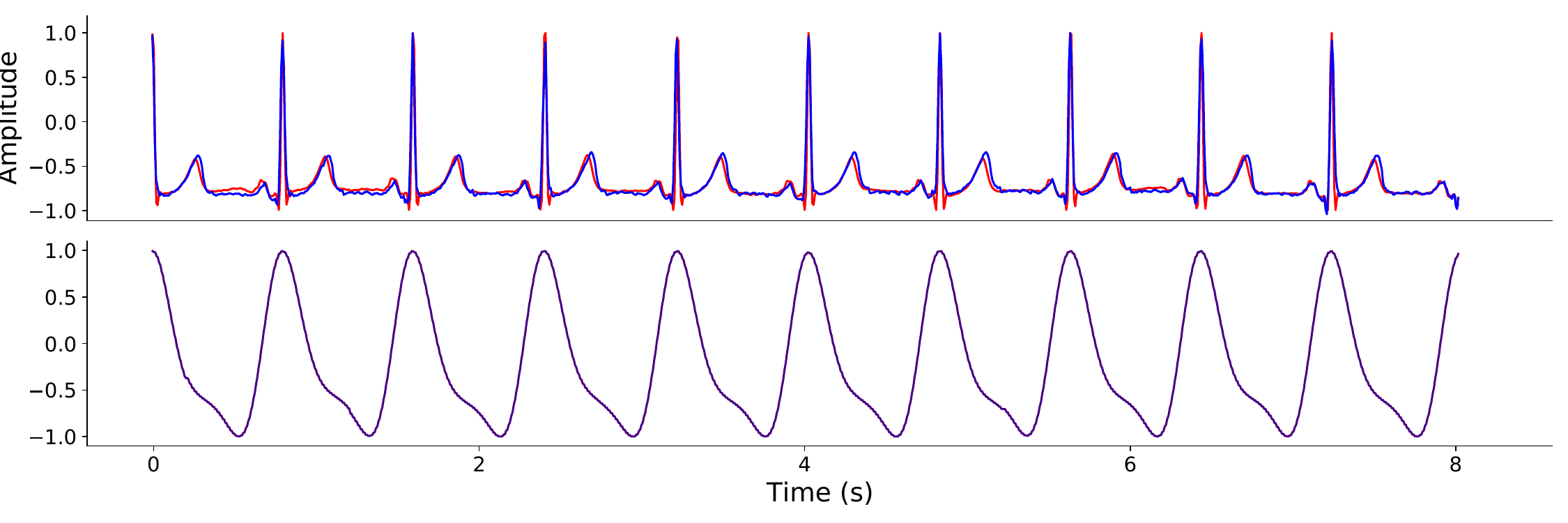}
    \caption{Record 0325.}
    \label{fig:record_0325} 
  \end{subfigure}
  \begin{subfigure}{0.4\textwidth}
    \centering
    \includegraphics[width=\linewidth]{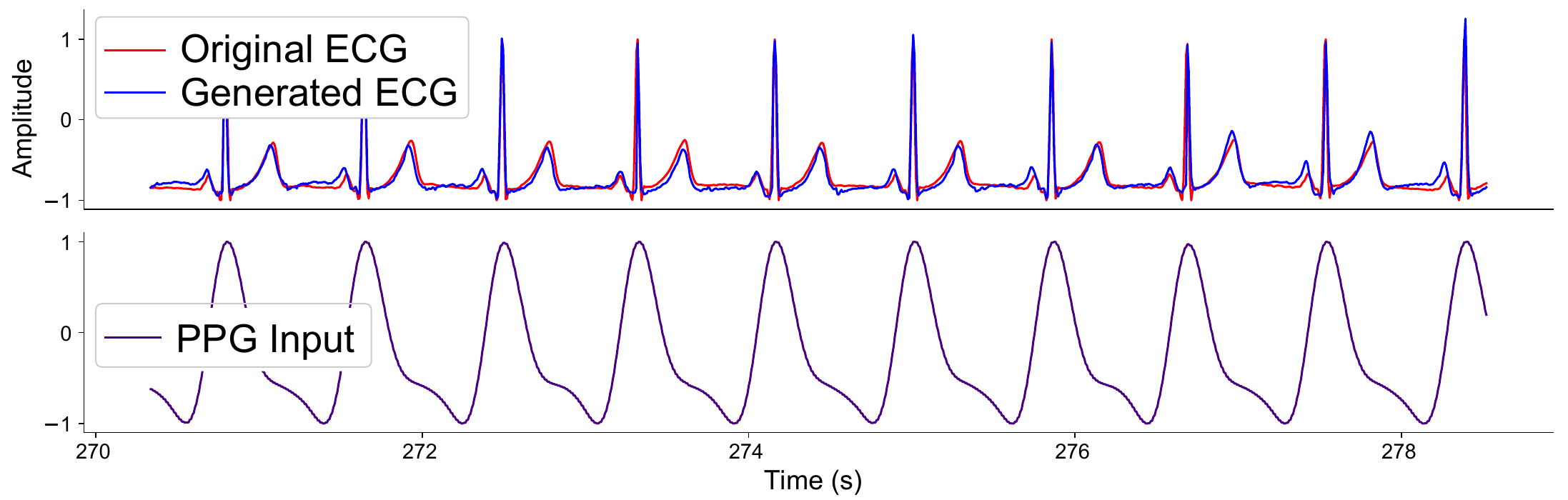}
    \caption{Record 0332.}
    \label{fig:record_0332} 
  \end{subfigure}
\caption{Visual comparison of ECG signals reconstructed by the proposed method using data from the CapnoBase dataset.}
\label{fig:record_capno_add_examples}
\vspace{-0.2cm}
\end{figure}

\begin{figure}[!ht]
\centering
  \begin{subfigure}{0.38\textwidth}
    \centering
    \includegraphics[width=\linewidth]{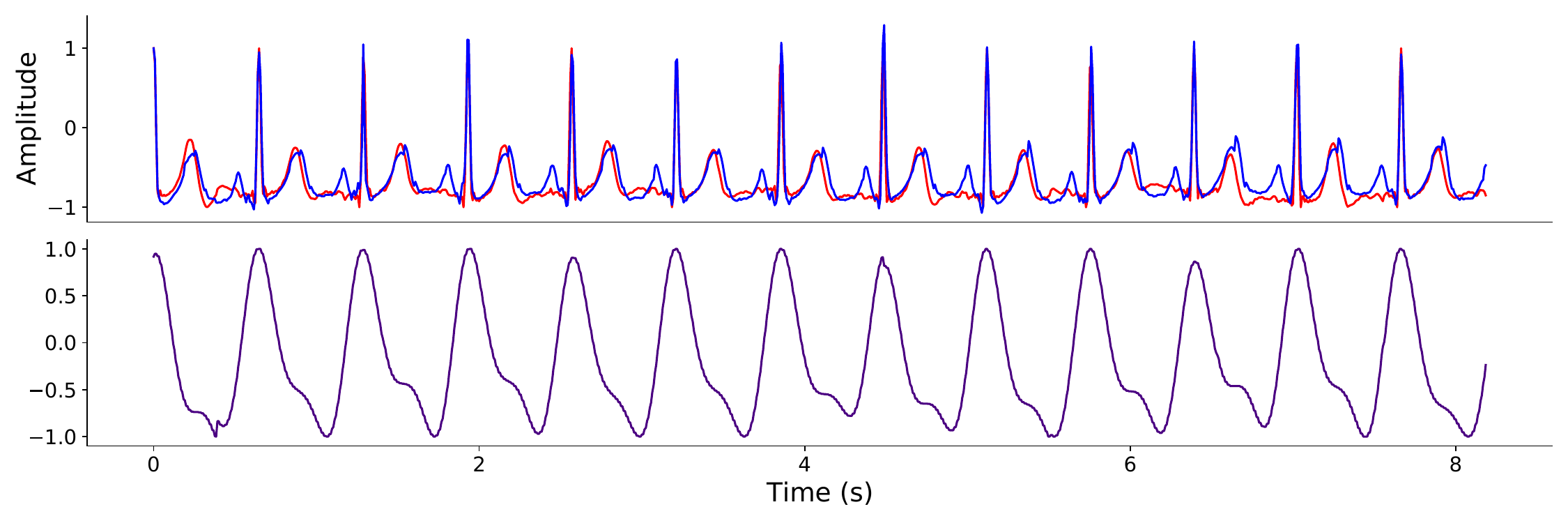}
    \caption{Record 01.}
    \label{fig:record_01} 
  \end{subfigure}
  \begin{subfigure}{0.38\textwidth}
    \centering
    \includegraphics[width=\linewidth]{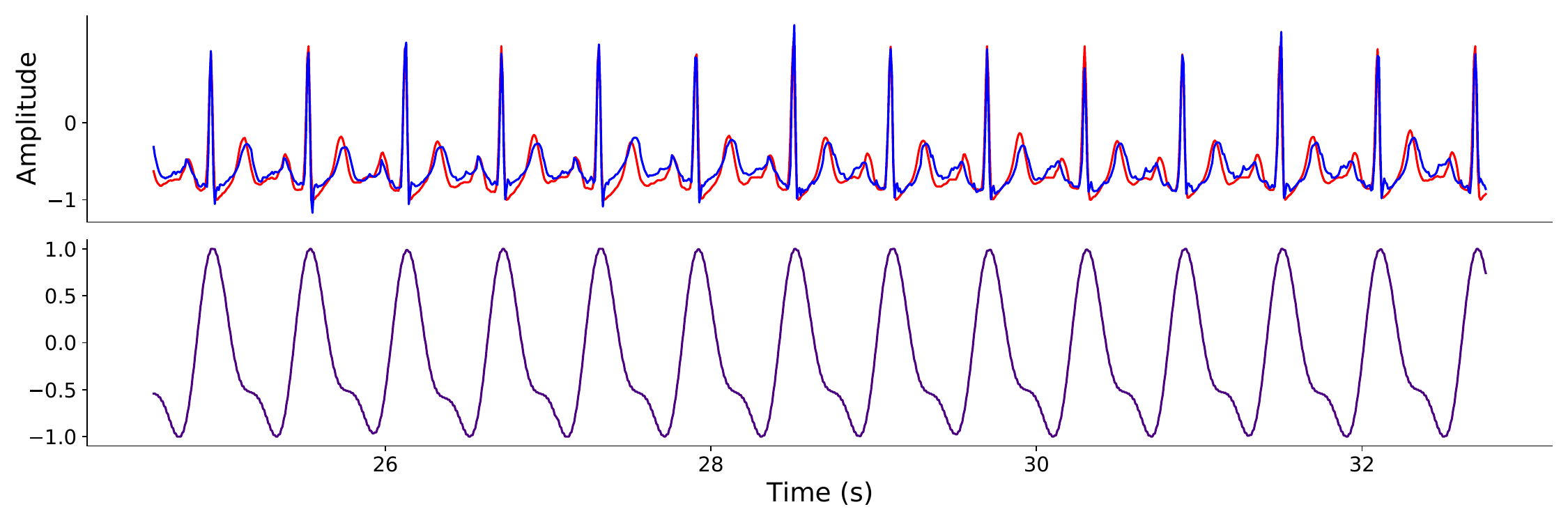}
    \caption{Record 08.}
    \label{fig:record_08} 
  \end{subfigure}
  \begin{subfigure}{0.38\textwidth}
    \centering
    \includegraphics[width=\linewidth]{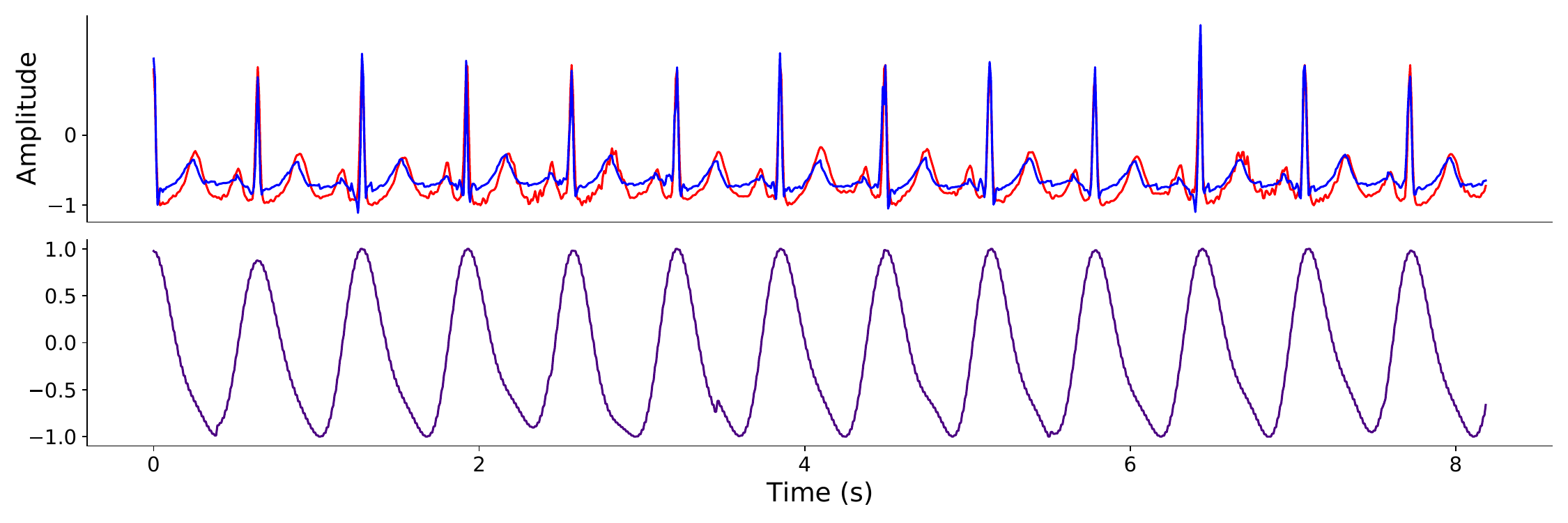}
    \caption{Record 14.}
    \label{fig:recored_14} 
  \end{subfigure}
  % \begin{subfigure}{0.3\textwidth}
  %   \centering
  %   \includegraphics[width=\linewidth]{./figs/add_examples/epoch2000_signal20_window12.pdf}
  %   \caption{Record 20.}
  %   \label{fig:record_20} 
  % \end{subfigure}
  \begin{subfigure}{0.38\textwidth}
    \centering
    \includegraphics[width=\linewidth]{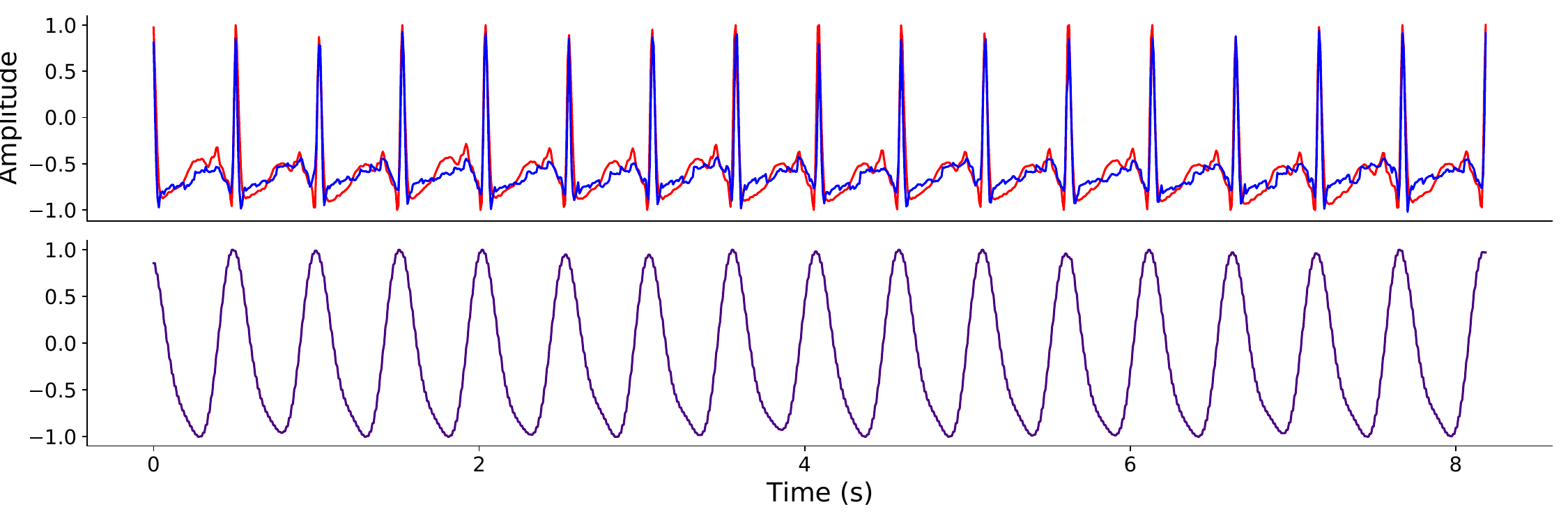}
    \caption{Record 21.}
    \label{fig:record_21} 
  \end{subfigure}
  \begin{subfigure}{0.38\textwidth}
    \centering
    \includegraphics[width=\linewidth]{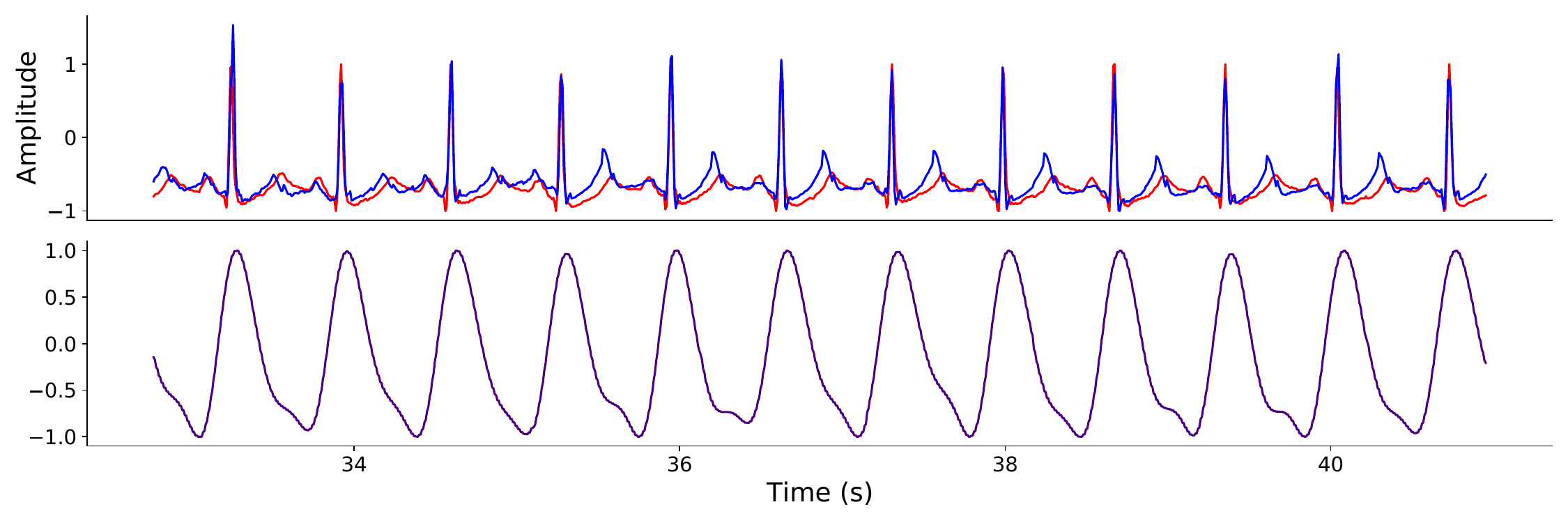}
    \caption{Record 22.}
    \label{fig:record_22} 
  \end{subfigure}
  \begin{subfigure}{0.38\textwidth}
    \centering
    \includegraphics[width=\linewidth]{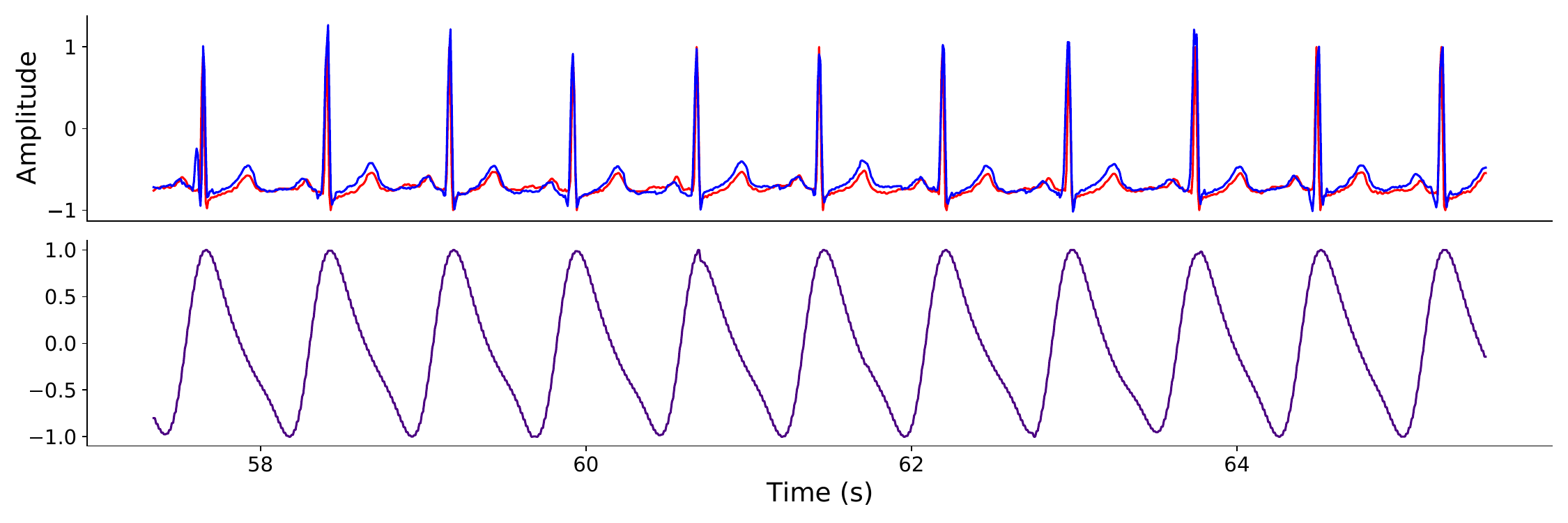}
    \caption{Record 34.}
    \label{fig:record_34} 
  \end{subfigure}
  \begin{subfigure}{0.38\textwidth}
    \centering
    \includegraphics[width=\linewidth]{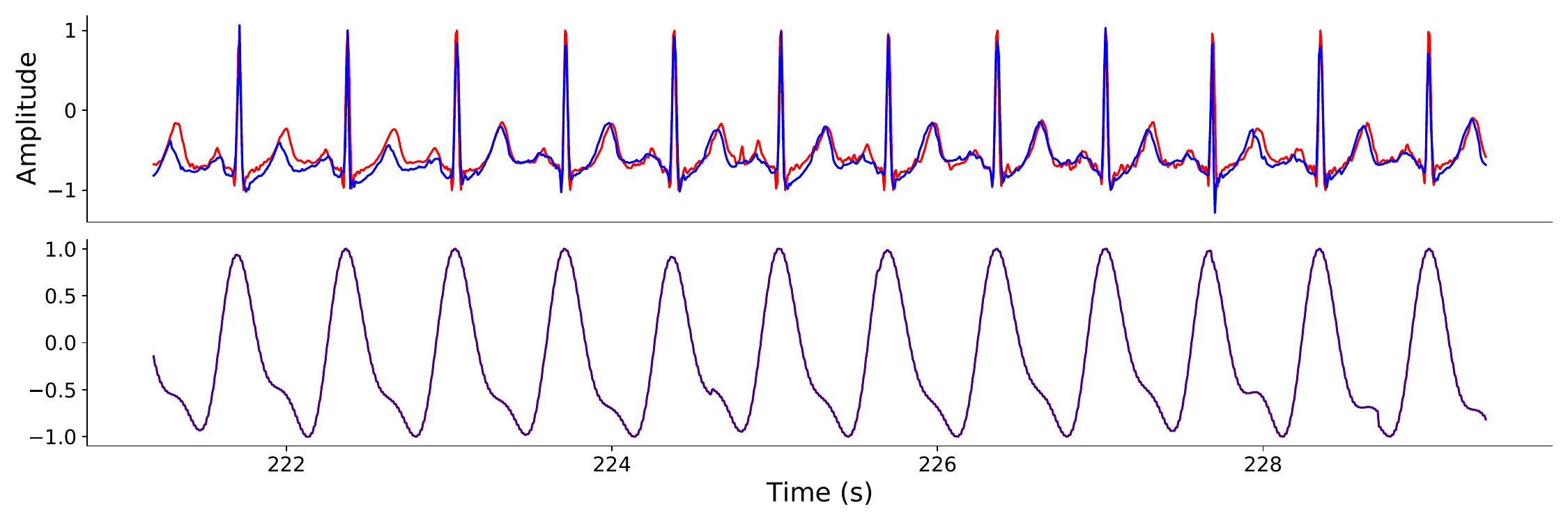}
    \caption{Record 37.}
    \label{fig:record_37} 
  \end{subfigure}
  \begin{subfigure}{0.38\textwidth}
    \centering
    \includegraphics[width=\linewidth]{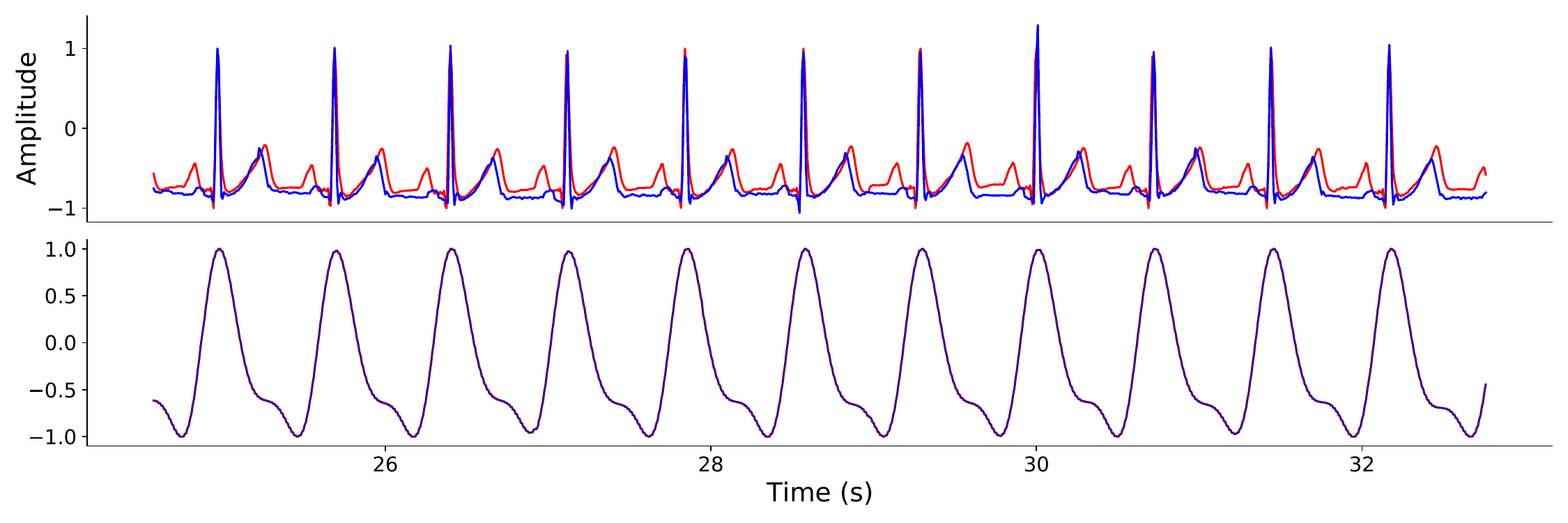}
    \caption{Record 42.}
    \label{fig:recored_42} 
  \end{subfigure}
  \begin{subfigure}{0.4\textwidth}
    \centering
    \includegraphics[width=\linewidth]{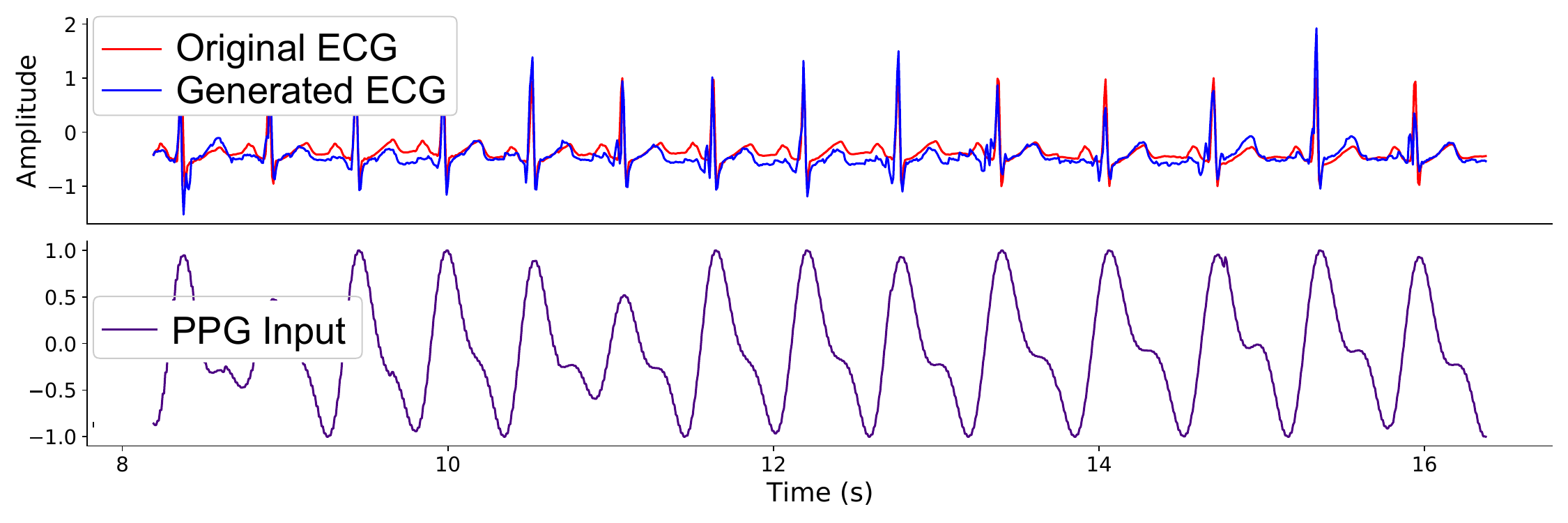}
    \caption{Record 48.}
    \label{fig:record_48} 
  \end{subfigure}
\caption{Visual comparison of ECG signals reconstructed by the proposed method using data from the BIDMC dataset.}
\label{fig:record_bidmc_add_examples}
\end{figure}

Figures~\ref{fig:record_capno_add_examples} and~\ref{fig:record_bidmc_add_examples} present additional visualizations of reconstructed ECG signals from two datasets. These examples demonstrate that our proposed method achieves promising results in PPG-to-ECG reconstruction, highlighting its potential for using PPG as an alternative measurement of heart activity.

 \bibliographystyle{elsarticle-num} 
 \bibliography{bibliography.bib}
\end{document}